\documentstyle[aps]{revtex}
\begin{document}
\draft
\input{epsf}
\title{Measurements of Light Nuclei Production in 11.5 A GeV/c Au~+~Pb 
Heavy-Ion Collisions}
\author{
T.A. Armstrong                \unskip,$^{(8,\ast)}$
K.N. Barish                   \unskip,$^{(3)}$
S. Batsouli                   \unskip,$^{(13)}$
S.J. Bennett                  \unskip,$^{(12)}$
M. Bertaina                   \unskip,$^{(7,\dag)}$
A. Chikanian                  \unskip,$^{(13)}$
S.D. Coe                      \unskip,$^{(13,\ddag)}$
T.M. Cormier                  \unskip,$^{(12)}$
R. Davies                     \unskip,$^{(9,\S)}$
C.B. Dover                    \unskip,$^{(1,\|)}$
P. Fachini                    \unskip,$^{(12)}$
B. Fadem                      \unskip,$^{(5)}$
L.E. Finch                    \unskip,$^{(13)}$
N.K. George                   \unskip,$^{(13)}$
S.V. Greene                   \unskip,$^{(11)}$
P. Haridas                    \unskip,$^{(7,\P)}$
J.C. Hill                     \unskip,$^{(5)}$
A.S. Hirsch                   \unskip,$^{(9)}$
R. Hoversten                  \unskip,$^{(5)}$
H.Z. Huang                    \unskip,$^{(2)}$
H. Jaradat                    \unskip,$^{(12)}$
B.S. Kumar                    \unskip,$^{(13,\ast\ast)}$
T. Lainis	              \unskip,$^{(10)}$
J.G. Lajoie                   \unskip,$^{(5)}$
R.A. Lewis                    \unskip,$^{(8)}$
Q. Li                         \unskip,$^{(12)}$
B. Libby                      \unskip,$^{(5,\dag\dag)}$
R.D. Majka                    \unskip,$^{(13)}$
T.E. Miller                   \unskip,$^{(11)}$
M.G. Munhoz                   \unskip,$^{(12)}$
J.L. Nagle                    \unskip,$^{(4)}$
I.A. Pless                    \unskip,$^{(7)}$
J.K. Pope                     \unskip,$^{(13,\ddag\ddag)}$
N.T. Porile                   \unskip,$^{(9)}$
C.A. Pruneau                  \unskip,$^{(12)}$
M.S.Z. Rabin                  \unskip,$^{(6)}$
J.D. Reid                     \unskip,$^{(11)}$
A. Rimai                      \unskip,$^{(9,\S\S)}$
A. Rose                       \unskip,$^{(11)}$
F.S. Rotondo                  \unskip,$^{(13,\|\|)}$
J. Sandweiss                  \unskip,$^{(13)}$
R.P. Scharenberg              \unskip,$^{(9)}$
A.J. Slaughter                \unskip,$^{(13)}$
G.A. Smith                    \unskip,$^{(8)}$
M.L. Tincknell                \unskip,$^{(9,\P\P)}$
W.S. Toothacker               \unskip,$^{(8)}$
G. Van Buren                  \unskip,$^{(2)}$
F.K. Wohn                     \unskip,$^{(5)}$
Z. Xu                         \unskip$^{(13)}$}
\address{\centerline{(The E864 Collaboration)}}
\address{  $^{(1)}$ Brookhaven National Laboratory, Upton, 
New York 11973 \break
  $^{(2)}$ University of California at Los Angeles, Los Angeles, 
California 90095 \break  
  $^{(3)}$ University of California at Riverside, Riverside, 
California 92521 \break
  $^{(4)}$ Columbia University, New York 10027 \break
  $^{(5)}$ Iowa State University, Ames, Iowa 50011 \break 
  $^{(6)}$ University of Massachusetts, Amherst, Massachusetts 01003 \break 
  $^{(7)}$ Massachusetts Institute of Technology, Cambridge, 
Massachusetts 02139 \break 
  $^{(8)}$ Pennsylvania State University, University Park, 
Pennsylvania 16802 \break 
  $^{(9)}$ Purdue University, West Lafayette, Indiana 47907 \break 
  $^{(10)}$ United States Military Academy, West Point, New York 10996 \break
  $^{(11)}$ Vanderbilt University, Nashville, Tennessee 37235 \break 
  $^{(12)}$ Wayne State University, Detroit, Michigan 48201 \break 
  $^{(13)}$ Yale University, New Haven, Connecticut 06520 \break
}

\date{\today}
\maketitle
\begin{abstract}
We report on measurements by the E864 experiment at the BNL-AGS
of the yields of light nuclei in collisions of 
$^{197}Au$ with beam momentum of 11.5 A GeV/c on targets of
$^{208}Pb$ and $^{197}Pt$.  The yields are reported for
nuclei with baryon number A=1 up to A=7, and typically cover
a rapidity range from $y_{cm}$ to $y_{cm}+1$ and a transverse momentum 
range of approximately
$0.1 \leq p_{T}/A \leq 0.5 $GeV/c.  We calculate coalescence
scale factors $B_{A}$ from which we extract model dependent source dimensions
and collective flow velocities.  We also examine the dependences of the 
yields on
baryon number, spin, and isospin of the produced nuclei.
\end{abstract}
\pacs{25.75.-q}

\section{Introduction}

Relativistic heavy ion collisions are believed to reach energy densities
an order of magnitude greater than that of normal nuclear matter.  
These collisions allow the examination of the strong
interaction in a novel environment as well as providing a possible
doorway to new states of matter.  In order to understand the dynamics of
the collision system, one must use the only available 
tools-the species and momenta
of the particles which exit the collision region.  The use of emitted
hadrons to probe the collision system is complicated because these
hadrons rescatter many times as they traverse the collision region, and
consequently lose some of their direct information about the earlier stages 
of the time evolution.  However, the final space-time extent of the system at
freeze-out (the time when strong interactions cease) and position-momentum
correlations of the emitted particles contain much information about
the entire time evolution.  In principle, these carry information about
the equation-of-state of the early collision region. 

In order to extract information about both the momentum and position
distributions of the source at freeze-out, it is
necessary to measure multi-particle correlations.  One widely
used technique is Hanbury-Brown-Twiss interferometry \cite{pratt}
for which the correlations between particles are due to quantum statistics.
Another method is through the measured yields of light nuclei,
which are formed by the coalescence of individual nucleons.

Because of the violence of heavy ion collisions, it is 
highly improbable for a nuclear cluster near center-of-mass
rapidity $y_{cm}$=1.6 in a collision at these energies to be a fragment of the
beam or target nucleus \cite{dq1and2} \cite{sjohn}.  
This would involve a cluster suffering a momentum loss
of several GeV/c per nucleon that does not destroy the
cluster, which is typically bound by only a few MeV 
per nucleon.  These nuclei then are formed by coalescence and so represent  
correlations of several nucleons.  As the mass of measured nuclei increases, of
course, so does the number of particles involved in
the correlation, and so does the sensitivity to features of the freeze-out
distribution.

In part due to the fragility of these states, the
observed light nuclei are believed to be formed only
near freeze-out of the collision system, at which
time the mean free path of a bound cluster is long
enough for it to escape without further collision.
It is this notion that gives rise to a class
of models of light nuclei production, the coalescence
models (for example, \cite{sato,bond,nagle}).  
In general, these models assume a phase space
distribution of nucleons at freeze-out and impose some
coalescence conditions on the freeze-out positions
and momenta of the nucleons in order to calculate the 
yields of nuclei.  These models differ both
in their assumptions about the phase space profiles
and coalescence conditions.  Their differences
are often characterized by their predictions of the
invariant coalescence, or $B_{A}$, parameters which are 
defined as 
\begin{equation}
B_{A} \equiv \frac{( E \frac{d^{3}N_{A}}{dP^{3}}) } {( E \frac{d^{3}N_{neutron}}
{dp^{3}})^{N} ( E \frac{d^{3}N_{proton}}{dp^{3}} )^{Z}}
\label{eq:ba}
\end{equation}
where a nucleus with baryon number $A$ and momentum $P = pA$
is formed out of $Z$ protons and $N$ neutrons.

In the simplest momentum space coalescence models, coalescence
is assumed to take place between any nucleons with a small
enough momentum difference.   Early experimental results at the Bevalac with 
beam energies
of $\approx  500 A $MeV and high energy proton induced reactions 
revealed $B_{A}$ parameters which were approximately 
constant for these different collision systems (\cite{na52ba,nagle} contain 
useful summaries).  The assumption of such simple models is that if the collision
spatial volume is similar to size of the cluster, all nucleons whose
momentum difference is less than a fixed value will fuse to form a
nuclear cluster.  Thus the experimentally observed constant $B_{A}$ values 
may indicate collision volumes in these systems that are 
not substantially larger than the RMS size of a deuteron.

In more advanced models, assuming a quantum
mechanical sudden approximation \cite{bond} and using
a density matrix formalism, accounting for both the positions
and momenta of the nucleons \cite{sato}, the $B_{A}$ parameters take
on a relationship with the source volume $V$ of
$B_{A} \propto (1/V)^{(A-1)}$.  Heavy-ion experimental results at higher
energies at the AGS ($\approx 10 A GeV$) and CERN ($\approx 160 A GeV$) 
revealed $B_{A}$ values that decreased with beam energy.  This
observation was understood to be a sign of significant expansion in the 
collision volume before freeze-out.  This larger source volume creates a 
situation where some nucleons with small relative momentum will have
too large a spatial separation to coalesce, thus reducing $B_{A}$.

The density matrix formalism \cite{sato} assumes that although
the collision volume can expand significantly, there is no
correlation between the momentum and the position of a given
particle.  This assumption then leads to a prediction 
of no kinematic dependence 
of the $B_{A}$ parameter.  However, there is a great deal of
evidence that collective motion is present leading to expansion of
the collision volume and significant position-momentum correlations
\cite{pbm} \cite{uliflow}.  
Although the overall expansion of the system tends to decrease $B_{A}$ values
by spatially isolating nucleons from each other, collective motion makes it
more likely that nucleons that are spatially close together also have
similar momenta, which to some extent works in the opposite direction
by increasing $B_{A}$.  Other coalescence models have made 
an effort to include the effects of both larger source volumes and 
collective motion, by including coalescence as a
'afterburner' in collision cascade models such as RQMD 
\cite{RQMD,nagle} and by analytical calculations \cite{heinzprc} \cite{polleri}.

Light nuclei production can also be calculated from thermal
models \cite{mek} \cite{pbmtherm,pbm} which assume 
at least local thermal equilibrium of
the system and thus that particle production (including,
in this case, composite particles) is governed
by a single temperature and chemical potentials.  This gives
rise to expressions for the $B_{A}$ with the 
same $(1/V)^{(A-1)}$ dependence found in some coalescence
models.  Collective expansion can be included in these models
; it affects only the amount of energy available
in local rest frames for particle production.

There has been much experimental effort in the measurement of light
nuclei in heavy ion reactions.
Previous measurements at AGS energies, including clusters of $A \leq 4$,
show values for $B_{A}$ that are considerably lower than 
at Bevalac energies \cite{nag81}, indicating a much greater 
expansion of the system.
This is also consistent with AGS results showing that 
the $B_{A}$ become smaller with more central 
collisions and larger target nuclei \cite{bennet} \cite{abb94}.
At CERN-SPS energies, production of secondary particles
(chiefly pions) is several times as large as at the AGS,
leading to a larger expansion before hadronic freeze-out.
Coalescence of high mass clusters is thus much less
probable, as indicated by yet lower
values for the $B_{A}$ as measured for example by experiment 
NA52 \cite{na52ba}.  

In this paper we describe and report the results of measurements
by E864 of the yields of light nuclei in collisions of 
$^{197}Au$ with beam momentum of 11.5 A GeV/c on targets of
$^{208}Pb$ and $^{197}Pt$.  The yields are measured for
nuclei from baryon number A=1-7.  In Section~\ref{sec-exp} 
we briefly describe the 
experimental apparatus and the analyses used to
produce our final invariant multiplicities.  In 
Section~\ref{sec-results} we report the results and
compare them with measurements of other
experiments where such measurements
overlap.  Finally in Section~\ref{sec-disc} we
examine trends in the data and discuss 
interpretations in the context of several different
models of light nuclei production.

\section{Experiment 864}
\label{sec-exp}

\subsection{Apparatus}
\begin{figure}
\centering\leavevmode\epsfbox{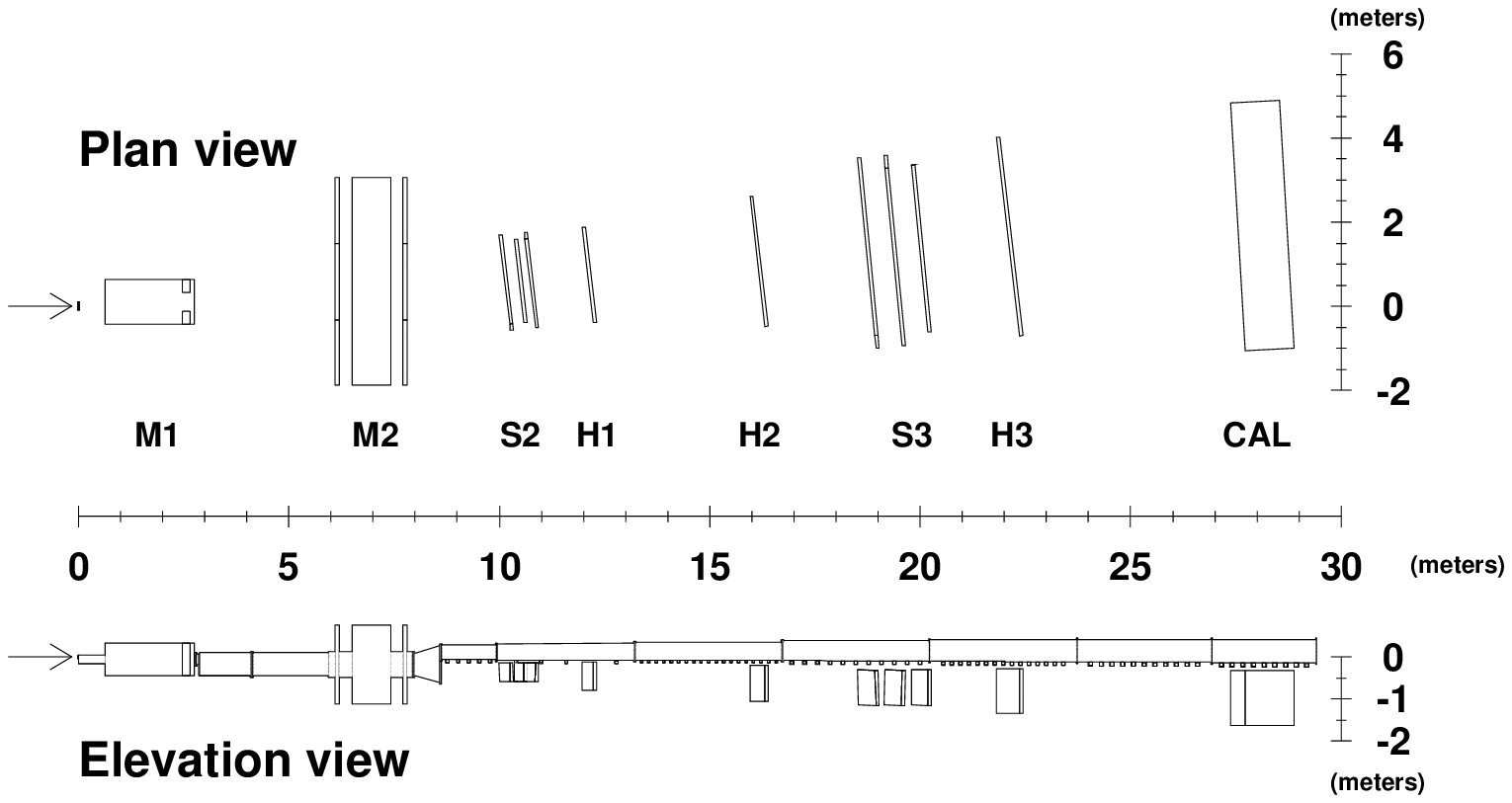}
\caption{The E864 spectrometer in plan and elevation views, showing the
dipole magnets (M1 and M2), hodoscopes (H1, H2, and H3), straw tube arrays 
(S2 and S3) and hadronic calorimeter (CAL).  The vacuum chamber is not 
shown in the plan view. }
\label{fig:john_ap}
\end{figure}

\begin{figure}
\centering\leavevmode\epsfbox[106 210 493 640]{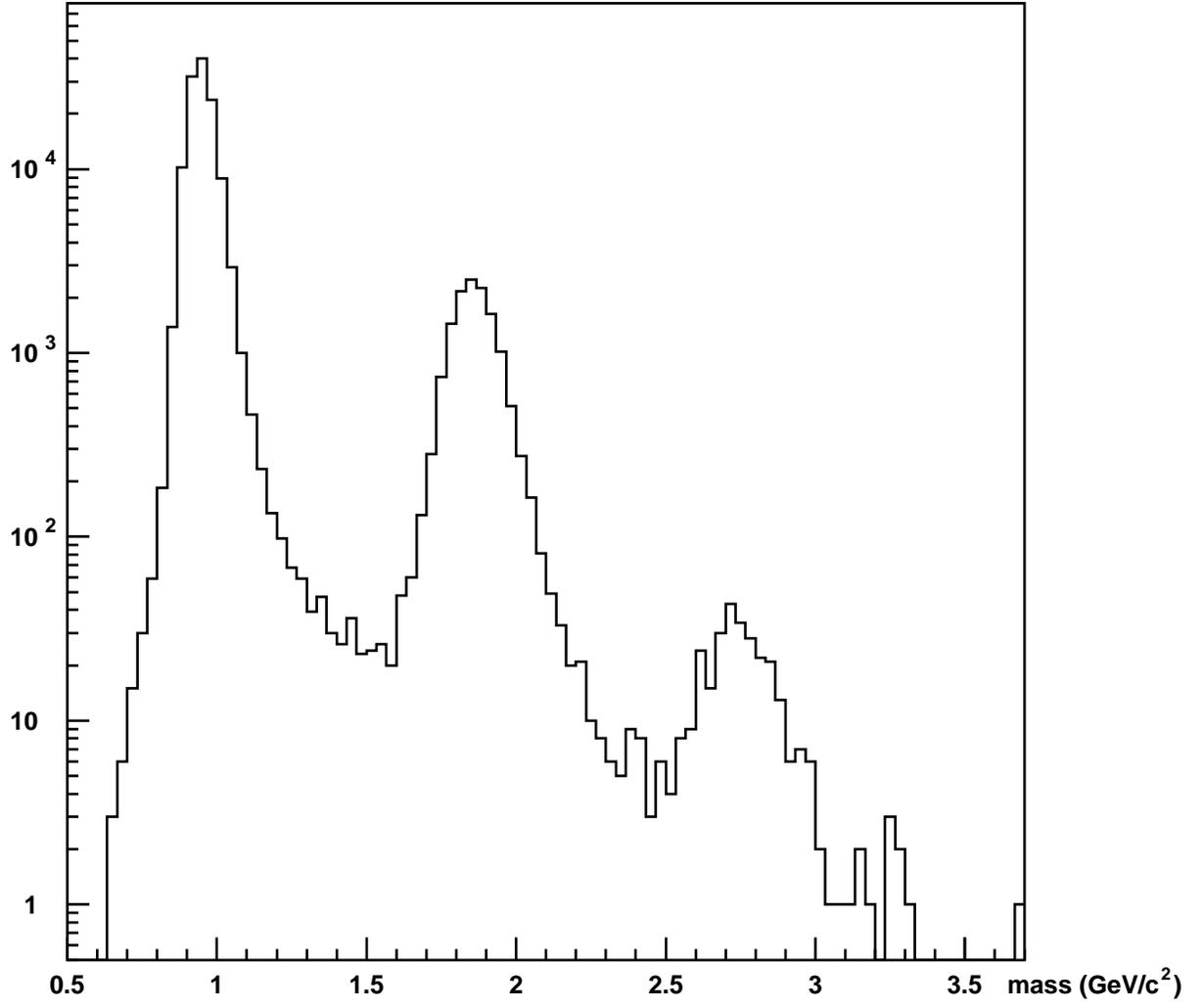}
\caption{Typical reconstructed mass spectrum for charge one
species in the rapidity slice $2.0 \leq y \leq 2.2 $ with a
magnetic field of .45 T.}
\label{fig:mass_dist}
\end{figure}

Brookhaven AGS Experiment 864 is an open geometry, high data
rate spectrometer which was chiefly designed to search for rarely 
produced objects in Au+Pb collisions.  Figure~\ref{fig:john_ap}  
shows a schematic
view of the experimental apparatus, a thorough description of which
is given in \cite{bignim}.  
Event centrality (impact parameter) is characterized by the charged 
particle multiplicity
measured in an annular scintillator counter \cite{beam} located 
approximately 10 cm downstream of the target, which subtends an
angular region from 16.6$^{o}$ to 45$^{o}$ in azimuth when viewed from 
the target.
The products of an interaction travel downstream through two dipole
spectrometer magnets, M1 and M2.  
Charged particle identification is performed
using information from the scintillator hodoscope walls (H1, H2, and H3) and
the straw tube stations (S2 and S3).  The hodoscopes walls each consist
of 206 1cm thick scintillator slats placed vertically. They provide information
about the charge, time-of-flight, and position of each charged particle
hit, and this information is used to identify candidate charged particle
tracks.  These tracks are then rejected or confirmed and further refined
by spatial hit information provided by the straw tube stations. 
Under the assumption that the track originates in the target, a rigidity is
assigned to the track by a look-up table generated from a GEANT simulation of
the experimental apparatus (using a technique described 
in \cite{alexi} as applied
for the PHENIX experiment at RHIC).  With information 
on rigidity, time-of-flight,
and charge, a mass can then be assigned to the track, 
providing particle identification.
A typical charge one mass distribution with a field of 0.45 Tesla in 
our spectrometer
magnets is shown in Figure~\ref{fig:mass_dist}.  Mass resolutions of 3 
to 7\% RMS are typical for particles with velocity $\beta \leq .985$.

At the downstream end of the apparatus is our hadronic calorimeter 
\cite{calo}; an array of 754 towers, each measuring 
10cm x 10cm on the front face, each of which 
provides energy and time information.  This
is the essential piece of the apparatus in our analyses of yields of neutral
particles, for which the tracking detectors serve only to provide a veto. 
In addition, the energy and timing 
information from each tower is used to provide
the input for a level II high mass trigger (the Late-Energy Trigger 
or LET \cite{let}), which provides an enhancement of 
approximately a factor of 50 
in our searches for rare high mass states.  

Measurements reported in this paper are from a variety of 
experimental conditions, including
different trigger conditions
and different magnetic field settings in M1 and M2.  The different 
data sets and their experimental
conditions are listed in Table~\ref{tab:datasum}.  Because of the 
large acceptance
open geometry design of the experiment, the different data sets often have
significant regions of overlap with one another, allowing a consistency
check on the measurements-see for example the measurements of alpha
particles introduced in Section~\ref{sec-results}.

\subsection{Data Analysis}

In order to measure the yield of a given species, a mass plot analogous to
Figure~\ref{fig:mass_dist} is made for each 
kinematic bin.  The number of tracks which lie
within its mass peak are determined.  Background 
under each peak is then estimated, 
generally with fits to signal plus background, and 
subtracted away from this count.  
The invariant multiplicity in a given kinematic
bin is then determined by correcting this 
number of raw counts for the geometric 
acceptance, trigger efficiency, charge cut 
efficiency, track quality cut efficiency,
and detector efficiency.  

\begin{figure}
\centering\leavevmode\epsfbox{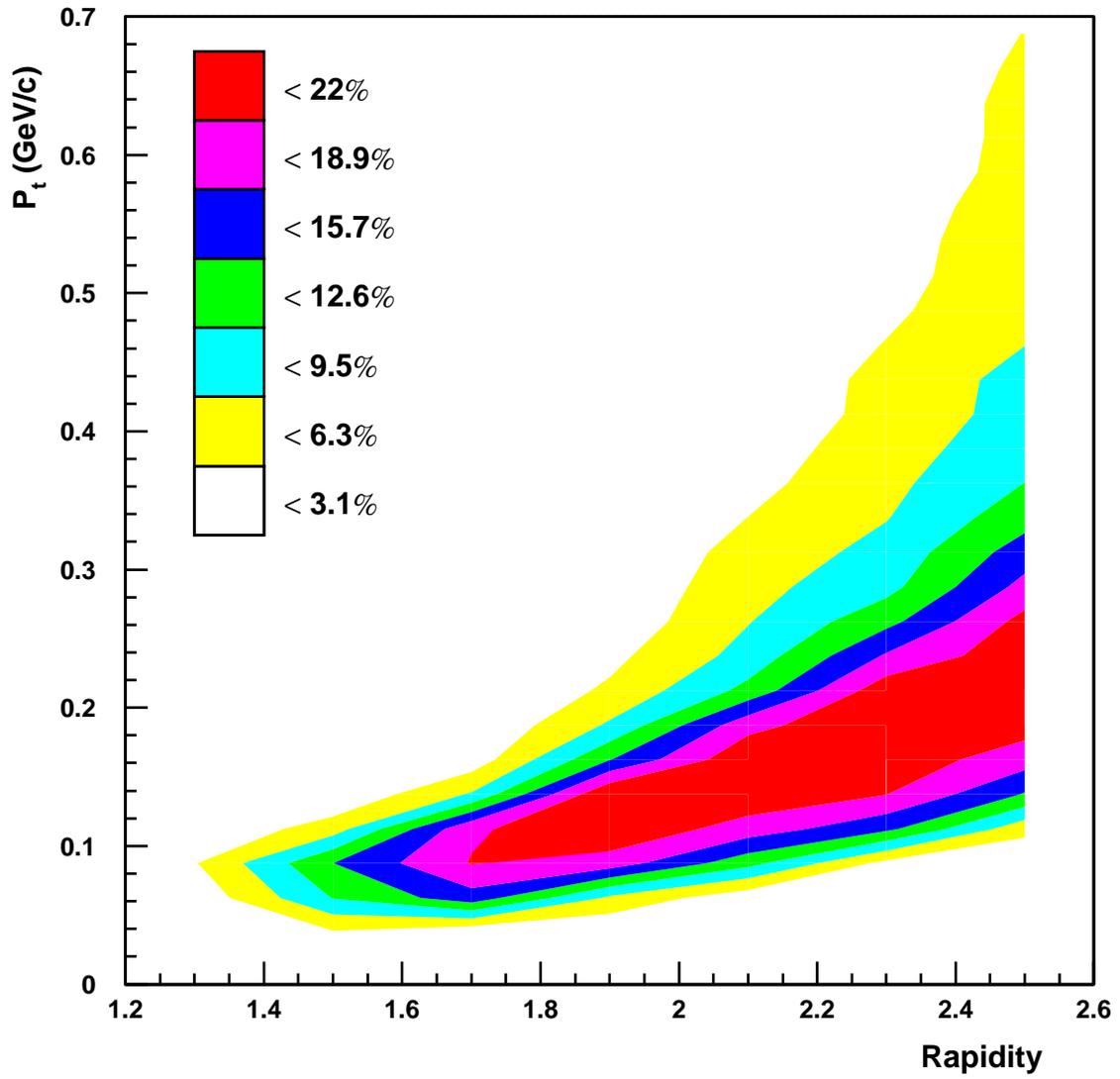}
\caption{Geometric acceptance efficiency for protons in the B=.45 T
field setting.}
\label{fig:pro_ax}
\end{figure}

Geometric acceptances are generally 25\% or lower as shown in 
Figure~\ref{fig:pro_ax} for protons.
Charge cut and track quality cut efficiencies 
are typically 90 to 95\%.  The three
redundant charge measurements for each track 
allow easy calculation of the charge cut efficiency
in each hodoscope simply by examining charge 
measurements made in each hodoscope 
against results from the other two; charge misidentification in each of the
three hodoscopes is less than one percent, so fewer than one track in a million
is assigned an incorrect charge.  In general, the track quality cuts are 
determined by comparison
with monte-carlo simulations.  In cases where the efficiencies
are particularly high, they are determined directly from the data.
The total track detector efficiency is 85 to 90\%; this is determined by
excluding each detector in turn from the track-finding process.  

The trigger efficiency is significant
only for those measurements which were made using the LET.  This mass and
momentum dependent efficiency ranges from approximately 
40\% to 90\% for the measurements reported here.  The LET
efficiency is determined by two methods.  The first,
used mainly for slow, high mass states for which the
efficiency is very high, is done simply using a monte-carlo 
simulation of the shower generated by the object and
knowledge of the LET look-up table in each tower.  The second
method is from the equation $\epsilon_{LET} = N_{LET}/(N_{LET}+R 
\times N_{nonLET})$ where R is the rejection factor provided by
the LET (i.e. number of events which fire the LET divided by
the total number of events).  $N_{LET}$ and $N_{nonLET}$
are the numbers of particles of interest which do and do not
produce LET triggers in LET triggered events, respectively.

Sources of systematic error that we have quantified include
possible error in the determination of the efficiencies listed
above as well as error in background subtraction (particularly
relevant for the deuterons and alpha particles at their
highest rapidities).  Also examined were the effects of changing
the assumed input distribution for each particle species in the
determination of geometrical acceptances and efficiencies and
effects of possible differences in the magnetic field with the
field maps that were used in reconstruction of tracks.  

Overall errors are generally dominated by systematics, particularly
for the lighter states.  Statistical errors can be significant
for the heavier states, particularly in the determination of 
LET efficiencies in which the number of particles of interest
which do not fire the LET generally has the largest statistical
error.   

\section{results}
\label{sec-results}

\begin{figure}
\centering\leavevmode\epsfbox[83 144 513 675]{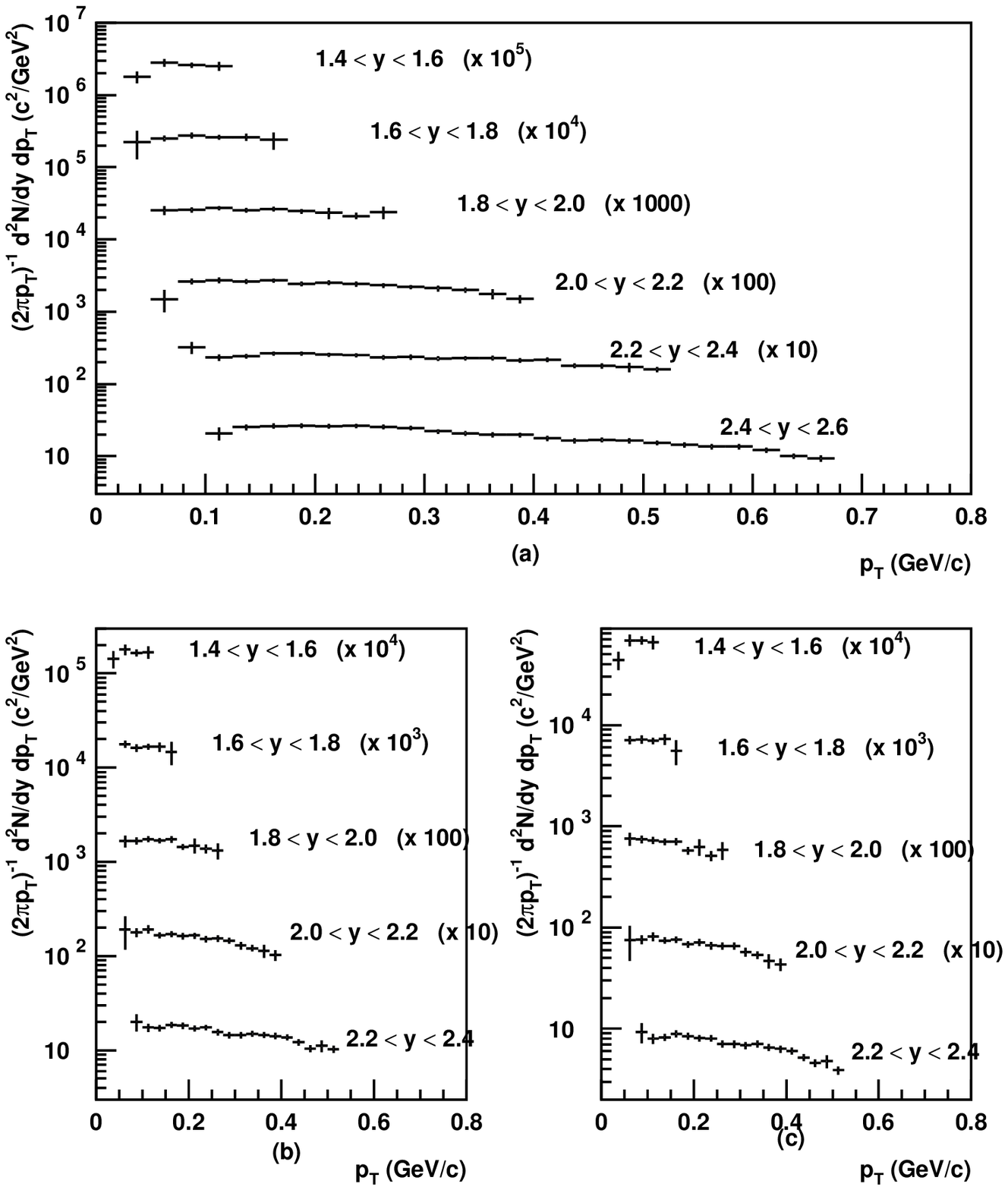}
\caption{Panel (a) displays invariant yields for protons in 10\% most
central Au+Pb collisions.  Panels (b) and (c) display the same
for 10-38\% and 38-66\% central collisions, respectively.  }
\label{fig:prot_mult}
\end{figure}

Measurements of invariant multiplicities 
for protons, neutrons, deuterons, $^{3}He$, 
and $^{4}He$ are shown in 
Figures~\ref{fig:prot_mult} through ~\ref{fig:alpha_mult},
and the values of the data shown in these figures
are listed in tables in the Appendix.
Figure~\ref{fig:prot_mult} displays proton 
invariant multiplicities for three different 
bins of collision centrality.  Figure~\ref{fig:inv_mult_np_2} 
shows only proton yields from 10\% most 
central Au+Pb collisions along with neutron multiplicities 
measured by E864 (see Reference \cite{neut}) for comparison.
Figures~\ref{fig:deut_mult} and~\ref{fig:he3_mult} 
display deuteron and $^{3}He$ invariant 
multiplicities for the same centrality bins used for the proton measurements.  
Tritons, $^{4}He$, $^{6}He$, $^{6}Li$, $^{7}Li$, 
and $^{7}Be$ are measured by E864 only in 10\% most 
central collisions; yields for tritons and 
alpha particles are shown in Figure~\ref{fig:trit_mult} and
Figure~\ref{fig:alpha_mult}, respectively, while yields for the heavier nuclei
are listed in tables in the Appendix.  
Added detail concerning most of
these measurements may be found in the Ph.D. theses 
listed in Reference \cite{theses}.

\subsection{Contributions from Hyperon Decays}
\label{sec-hyperons}

For comparison to other experimental results and calculations of 
coalescence parameters, it is
important to quantify the contribution to proton yields
that is made by protons which come from decays of hyperon
states, which to E864 are indistinguishable from primordial
protons.  There are three dominant hyperon decays which produce protons: 
$\Lambda    \rightarrow p + \pi ^{-}$, 
$\Sigma^{0} \rightarrow \Lambda + \gamma \rightarrow p + 
\pi ^{-} + \gamma$, and
$\Sigma^{+} \rightarrow p + \pi ^{0}$. 
The contributions from these decays were evaluated using a GEANT
simulation of the experiment with an input distribution taken
from measurements of E891 for the Lambda \cite{e891} and an input
distribution from RQMDv2.3 \cite{RQMD} for the Sigma.  From this simulation 
it was determined that protons from hyperon decays account for approximately
12\% of the measured yields of protons with only a slight kinematic 
dependence.  The proton and neutron 
yields from E864 shown in Figures~\ref{fig:prot_mult} 
and~\ref{fig:inv_mult_np_2}
have not been corrected for hyperon feed-down (nor have the values
listed in the appendix tables).

\subsection{Comparisons with Other Experimental Results}
\label{sec-comp}

\begin{figure}
\centering\leavevmode\epsfbox[101 168 511 671]{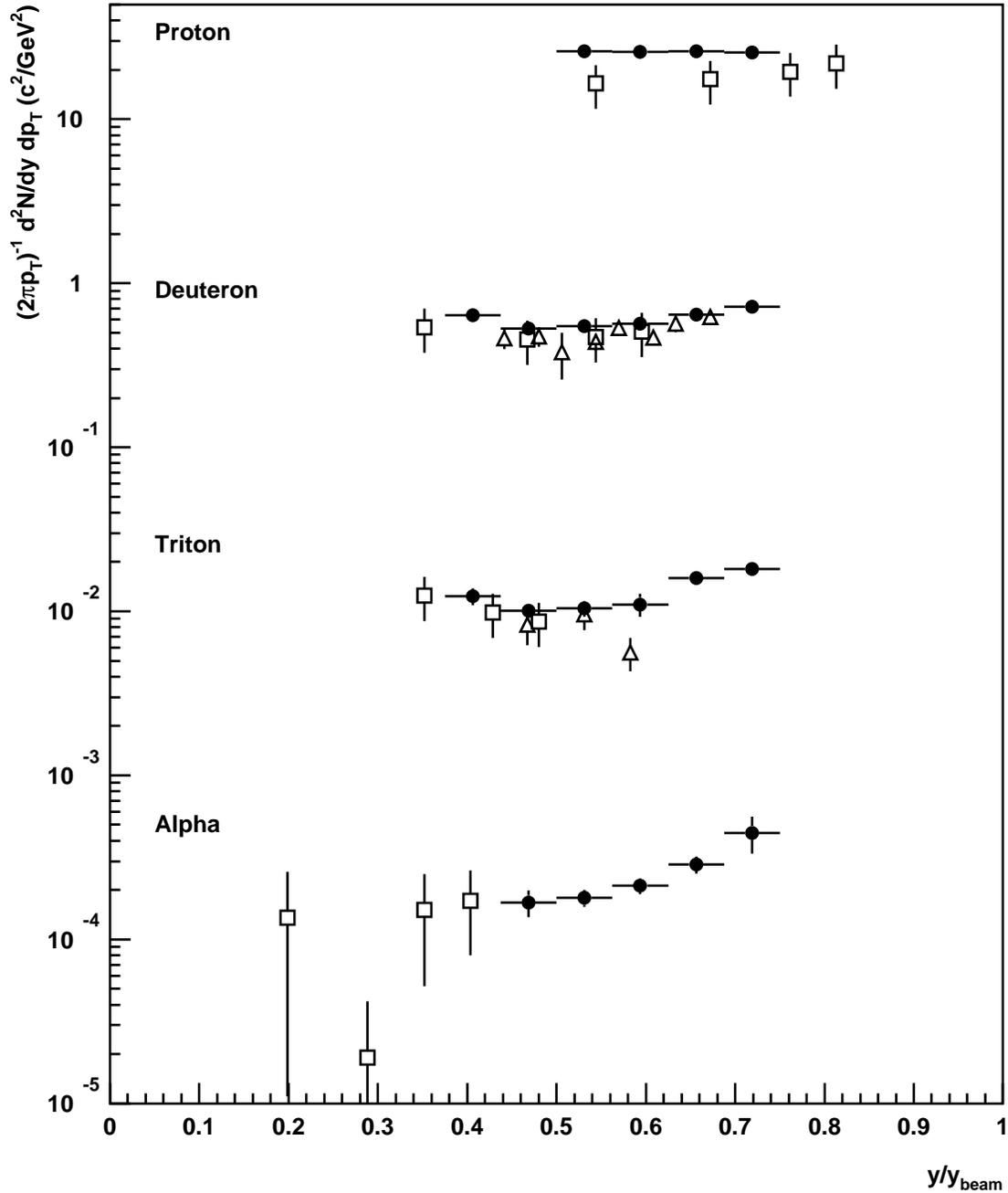}
\caption{Comparison of light nuclei yields measured by E864, E877 and 
E878.  The E864 results (solid circles) are the average of the yields 
over the transverse momentum range $.1 \leq \frac{p_{T}}{A} \leq .2$ 
GeV/c.  The E878 results (hollow squares) are the yields measured at
$p_{T} \simeq 0$.  The E877 points (hollow triangles) are measurements
at $.15 \leq \frac{p_{T}}{A} \leq .16$ GeV/c for deuterons and
$.11 \leq \frac{p_{T}}{A} \leq .22$ for tritons.  E864 and E878 points
represent yields in 10\% most central collisions while E877 points 
are from 4\% most central collisions. 
The E878 error bars are total errors including systematic errors except
for the alpha particle yields which contain an additional 25\% systematic
error which is not shown.  E864 error bars include both statistical and
systematic errors.}  
\label{fig:comp_e877_e878_paper}
\end{figure}

Figure~\ref{fig:comp_e877_e878_paper} shows a comparison of the 
light nuclei measurements from AGS experiments E864, E877 \cite{e877}, and
E878 \cite{bennet}.  Because of the different beam 
momenta of the experiments (10.8 A GeV/c in E878), the yields are plotted 
versus beam normalized rapidity.  The yields shown for E864 and E877 are 
average yields at approximately
$p_{T}/A$ = 150 MeV/c, while the E878 yields are measurements at 
$p_{T} \simeq 0$.  Other caveats to the comparisons
of the yields shown in Figure~\ref{fig:comp_e877_e878_paper} are noted
in the figure caption.

Proton yields measured by E864 are clearly higher than measurements
by E878.  Some of this difference can be attributed to protons which are
feed-down from hyperon decay.  The acceptance of E878 for these feed-down
protons is only about 10\% of what it is for primordial protons, while in
E864 the two acceptances are nearly the same.  When this difference is
taken into account (see Section~\ref{sec-hyperons}) and the 
E864 yields of primordial protons
are lowered by approximately 12\%, the results of the two experiments 
are different by approximately 25\% at midrapidity; this lies
within the range of systematic errors for the two experiments.  At
higher rapidities, the agreement is better.  

Comparison of the three experiments' measurements of deuterons and
tritons shows close agreement and the measurements of E864 and E878 of
alpha particle yields also agree within errors.

\section{Discussion}
\label{sec-disc}

\subsection{General Trends of the Spectra}
\label{sec-gen_tren}
\subsubsection{Transverse Dependence of Yields}

\begin{figure}
\centering\leavevmode\epsfbox[25 158 529 656]{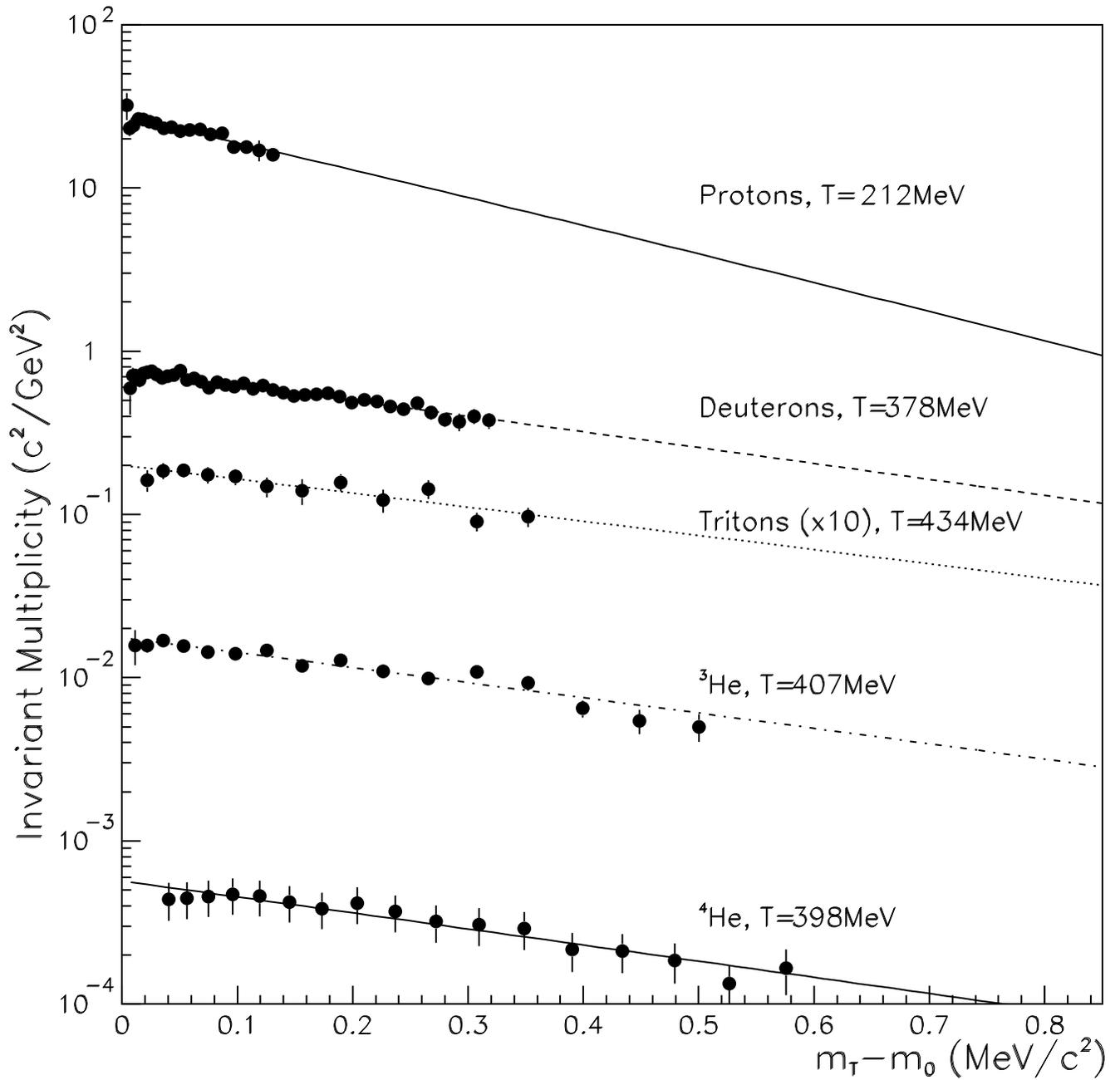}
\caption{Invariant yields of protons, deuterons, tritons, 
$^{3}He$ and $^{4}He$ in the rapidity range $2.2 \leq
y \leq 2.4$. Shown overlayed on each species is a fit to
the spectrum assuming a Boltzmann distribution in transverse
mass, with the extracted effective temperatures as noted.}
\label{fig:boltz_fits}
\end{figure}

E864 has sufficient coverage in transverse momentum for us to extract
measurements of inverse slope parameters of the 
different light nuclei yields in the rapidity range $2.2 \leq y \leq 2.4$.
Shown in Figure~\ref{fig:boltz_fits} are yields of protons,
deuterons, $^{3}He$, and $^{4}He$ as a function 
of transverse mass $m_{T} - m_{0} = \sqrt{p_{T}^{2}+
m_{0}^{2}} - m_{0} $ in this rapidity slice.  
Overlayed on the measurements are fits of
each species to a Boltzmann distribution in transverse mass, 
from each of which we 
extract an inverse slope parameter, $T$, as noted in 
Figure~\ref{fig:boltz_fits}.
The fits from which we extract $T$ are linear fits to the
log of the invariant multiplicities divided by $m_{T}$; this
is the same fitting method used for determining the slope parameters 
for neutrons in \cite{neut}.  The $\chi ^{2}$ values are
less than one per degree of freedom for all these fits.
%
%

\begin{figure}
\centering\leavevmode\epsfbox[16 165 529 660]{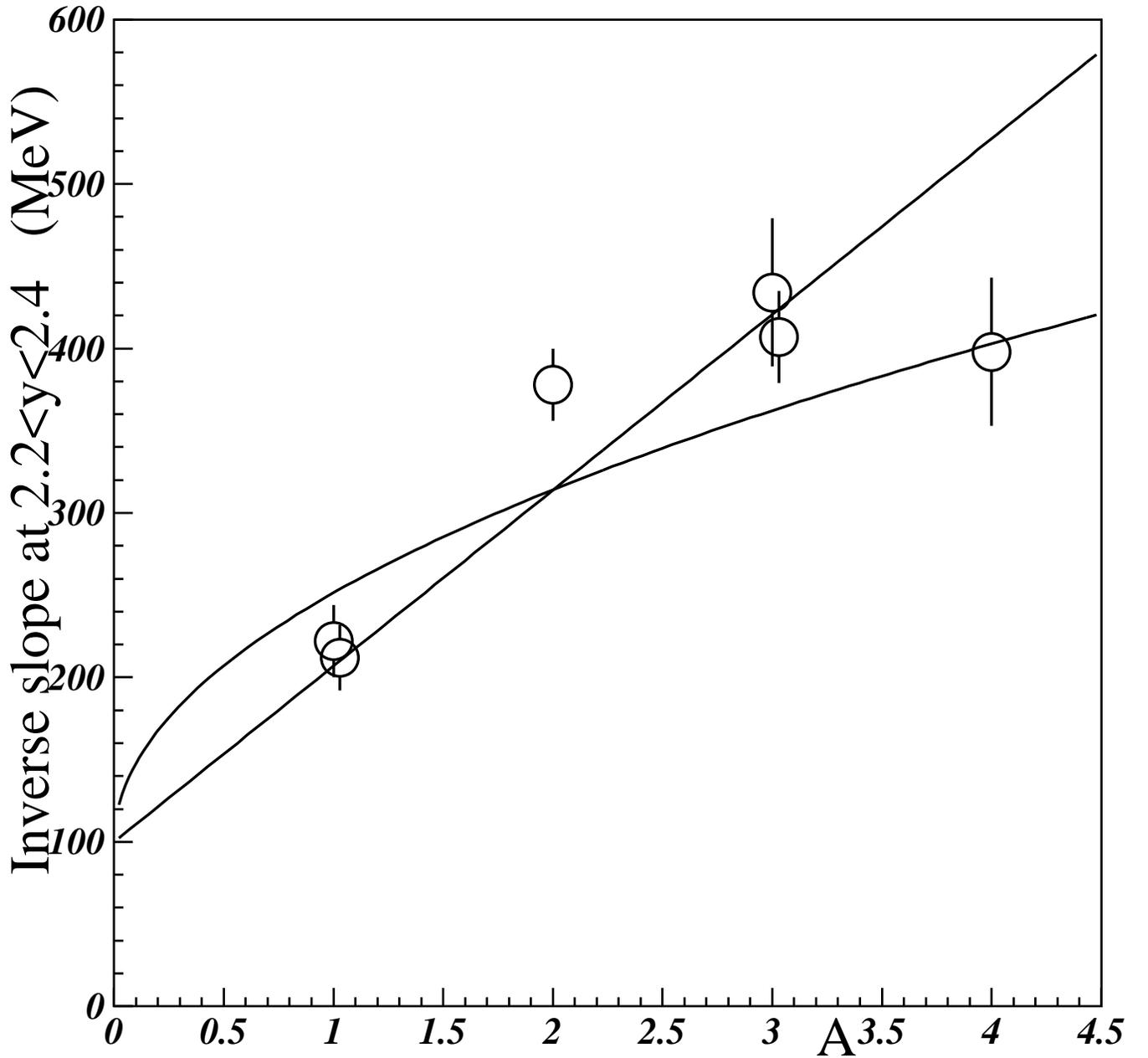}
\caption{Inverse slope parameters in the rapidity bin $2.2 \leq
y \leq 2.4$ shown as a function of mass number for protons, neutrons,
deuterons, tritons, $^{3}He$, and $^{4}He$.  Overlayed on the
points are curves generated by assuming (i) a box density profile which
gives rise to the straight line and (ii) a Gaussian profile with
velocity profile $v_{\perp} \propto (r/R)^{(1/2)}$. }
\label{fig:tvsm1}
\end{figure}

These slope parameters, as well as those for neutrons in this same 
rapidity range, are displayed 
in Figure~\ref{fig:tvsm1} as a function of mass number.  Polleri 
et. al. \cite{polleri} have demonstrated the sensitivity of these
trends in the inverse slopes to the density and velocity profiles of the 
nucleons
at the time when coalescence occurs.  To make this point, they have 
performed calculations
of the behaviour of these trends for different assumptions about the source
distributions.   Two of these assumptions give rise to the two curves 
shown overlayed on the data in Figure~\ref{fig:tvsm1}.  
The first has a 'box' spatial 
profile and a linear velocity profile, and the second has a 
Gaussian spatial distribution and a
velocity profile $v(r_{T}) \propto r_{T}^{(1/2)}$.  In determining
the curves shown in Figure~\ref{fig:tvsm1} we have fit the
data to these two different functional forms with the numerical
constraint of T=100 MeV for zero mass.  Neither of these sets of model
parameters provides an adequate description of the data.  
These curves are meant only to illustrate the sensitivity of 
these measurements: 
clearly, the sensitivity to the differences in these 
assumptions increases with  
increasing mass of the measured nuclei.  

\begin{figure}
\centering\leavevmode\epsfbox[16 165 529 660]{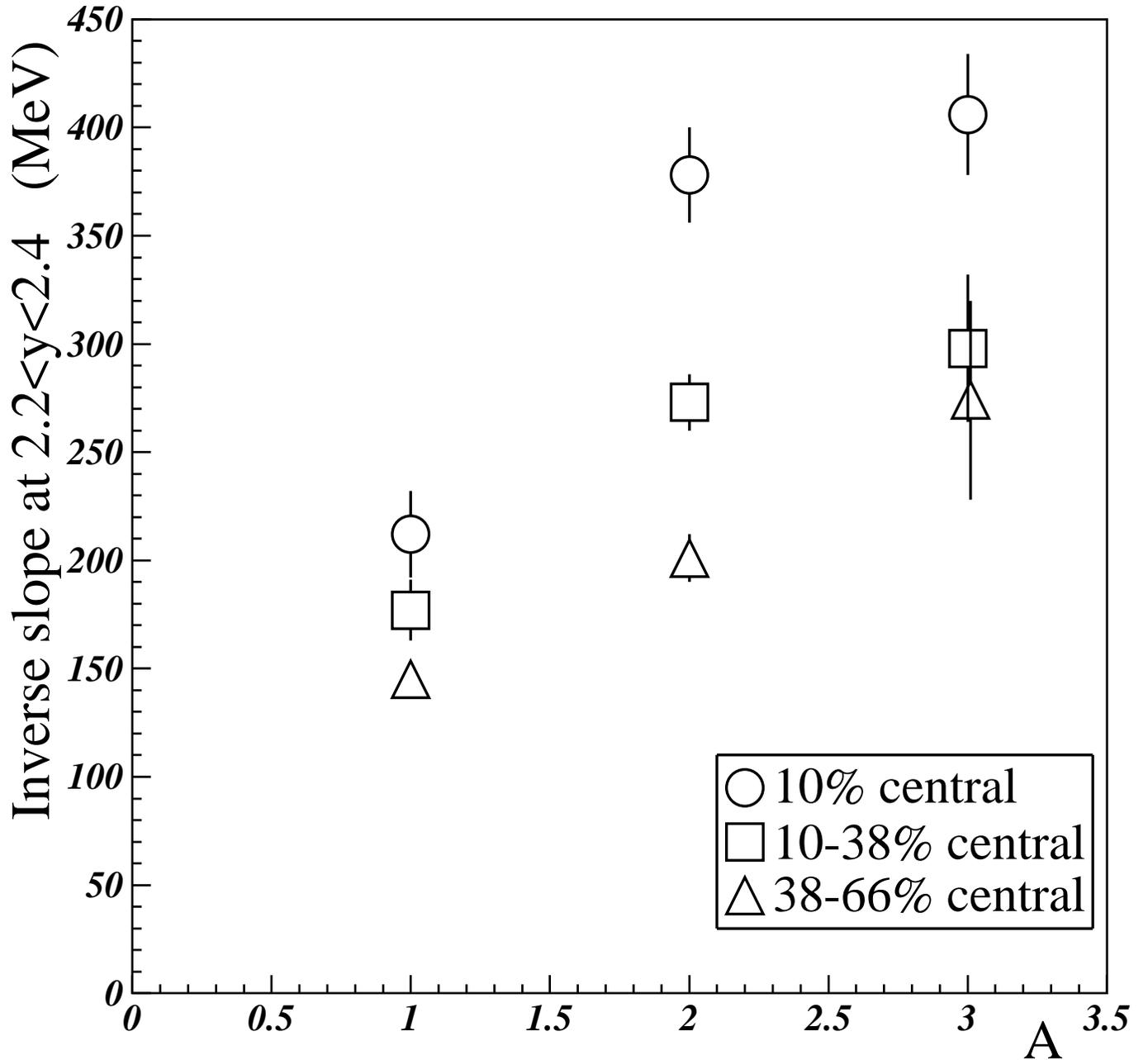}
\caption{Inverse slope parameters for protons, deuterons, and
$^{3}He$ in the rapidity bin $2.2 \leq
y \leq 2.4$ for three different collisions centralities.}
\label{fig:tvsm}
\end{figure}

Shown in Figure~\ref{fig:tvsm} are these same trends in light 
nuclei slope parameters including
centralities other than the 10\% most central collisions.  
For the more peripheral
events, the trends are consistent with a linear dependence of slope parameter on
mass (matching the calculation in Reference \cite{polleri} including box density
profiles).  Only for the most central events is there a clear rollover
in the slope as a function of mass.

\subsubsection{Longitudinal Dependence of Yields}

\begin{figure}
\centering\leavevmode\epsfbox[83 420 510 647]{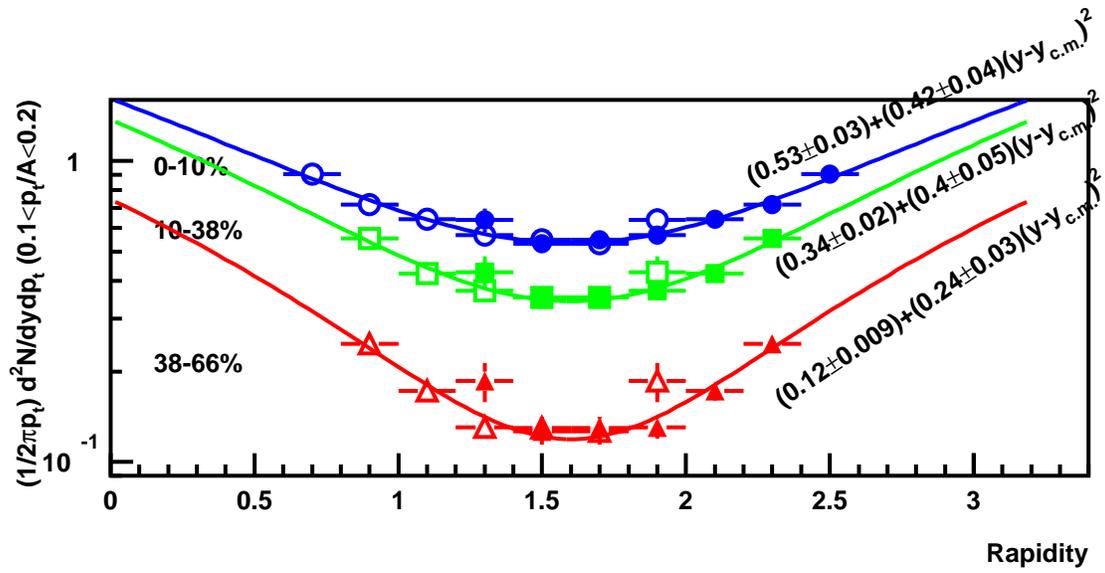}
\caption{Invariant yields of deuterons in the range of transverse
momentum $100 \leq \frac{p_{T}}{A} \leq 200$ MeV/c as a function
of rapidity for three different event centralities.}
\label{fig:conc_deut}
\end{figure}

In order to examine the rapidity dependence of the yields 
of light nuclei, we observe the
trends in multiplicities in the $p_{T}/A$ range from 100 to 200 MeV/c.  
(Our transverse coverage is not sufficient to integrate in transverse 
momentum for a 
full measurement of dN/dy over this entire rapidity range.)  
In Figure~\ref{fig:conc_deut} we show the invariant multiplicities of deuterons
for this low $p_{T}$ range for three different centralities as a 
function of rapidity.
The yields are concave as a function of rapidity (i.e. they are lowest
at center-of-mass and increase toward beam or target rapidity) and
become more concave for the more peripheral events.  We can parameterize
this concavity by fitting each set of yields to a quadratic
$a + b(y-y_{cm})^{2}$.  These fits are shown overlayed on the
data in Figure~\ref{fig:conc_deut}.  The ratio of coefficients, $b/a$,
from this parameterization serves as a measure of the relative
concavity of a species' spectrum as a function of rapidity at
low $p_{T}$ \cite{nkg_parkcity}.

\begin{figure}
\centering\leavevmode\epsfbox[82 209 504 633]{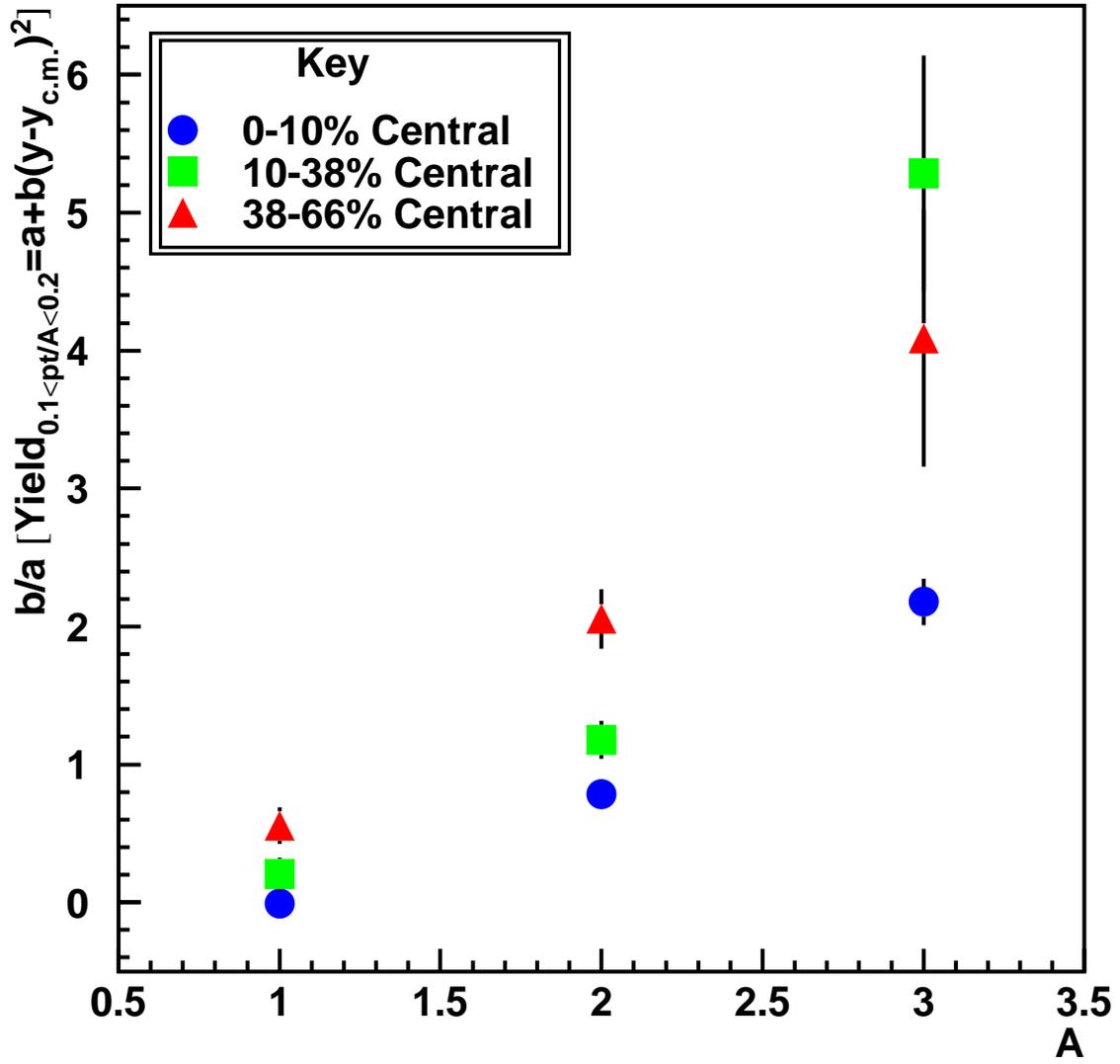}
\caption{Relative concavity of light nuclei spectra as 
a function of rapidity plotted versus mass number $A$ for
three different collision centralities.}
\label{fig:conc_mass}
\end{figure}

Values of this ratio $b/a$ are plotted in Figure~\ref{fig:conc_mass} 
as a function of mass number $A$ for protons, deuterons and $^{3}He$ 
in three different collision centralities.
We observe that the concavity of a spectrum increases with the mass of
a species and with collision centrality.  This can be understood
as some form of collective motion in the longitudinal direction, 
either expansion or incomplete stopping, which
increases in peripheral events.
It also is consistent with a rapidity dependent transverse expansion which
pushes the nuclei out of the transverse momentum range measured here.

\subsection{$B_{A}$ parameters}
\label{sec-ba}

The trends noted in Section ~\ref{sec-gen_tren} will also be observable
in a study of the behaviour of the $B_{A}$ parameters.
For simplicity, we can from our neutron measurements \cite{neut}  characterize
the neutron spectrum as a factor of 1.19$\pm$.08 
greater than the proton spectrum (feed-down
from hyperons is taken into account in determining this ratio), and so 
we evaluate $B_{A}$ as
\begin{equation}
B_{A} \equiv \frac{( E \frac{d^{3}N_{A}}{dP^{3}}) } 
{(1.19)^{N}( E \frac{d^{3}N_{proton}}
{dp^{3}})^{A}}.
\end{equation}
for a nucleus of baryon number $A$ with $N$ neutrons and $Z$ protons.  Again,
all invariant yields are evaluated at a common velocity.

\begin{figure}
\centering\leavevmode\epsfbox[59 399 532 681]{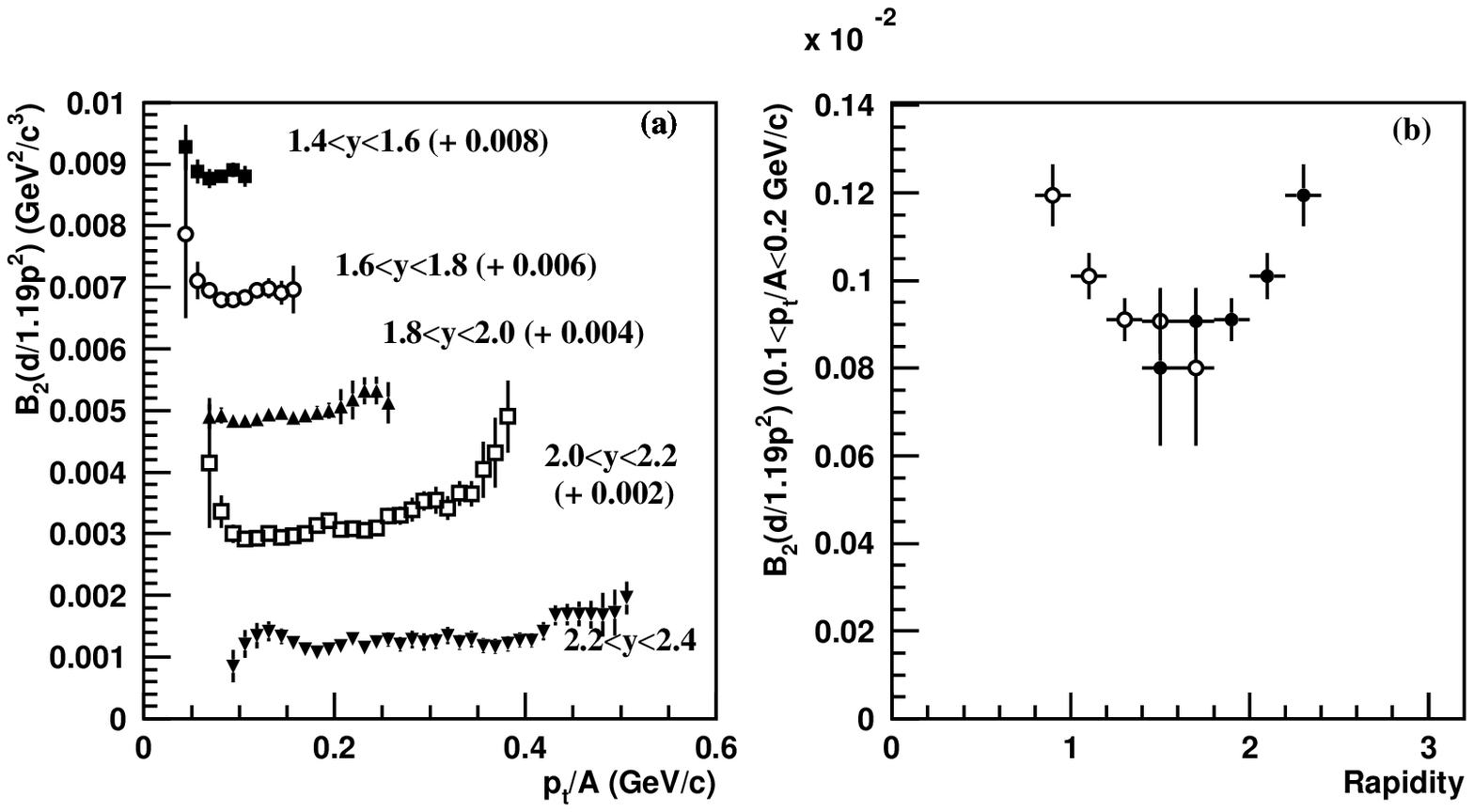}
\caption{Coalescence $B_{2}$ parameters, with an assumed neutron to
proton ratio of 1.19, shown (a) 
as a function of transverse momentum 
in several rapidity slices and (b)
as a function of rapidity near $p_{T}=0$.  
In panel (a), error bars include only point-to-point errors
(a rapidity dependent error of 5\% and a global 6\% error are not
included).  In panel (b), hollow points represent reflections 
about center-of-mass rapidity.  Here, the rapidity-dependent errors 
are included but the 6\% global systematic error is not.}
\label{fig:b2evan}
\end{figure}

\begin{figure}
\centering\leavevmode\epsfbox[57 408 526 680]{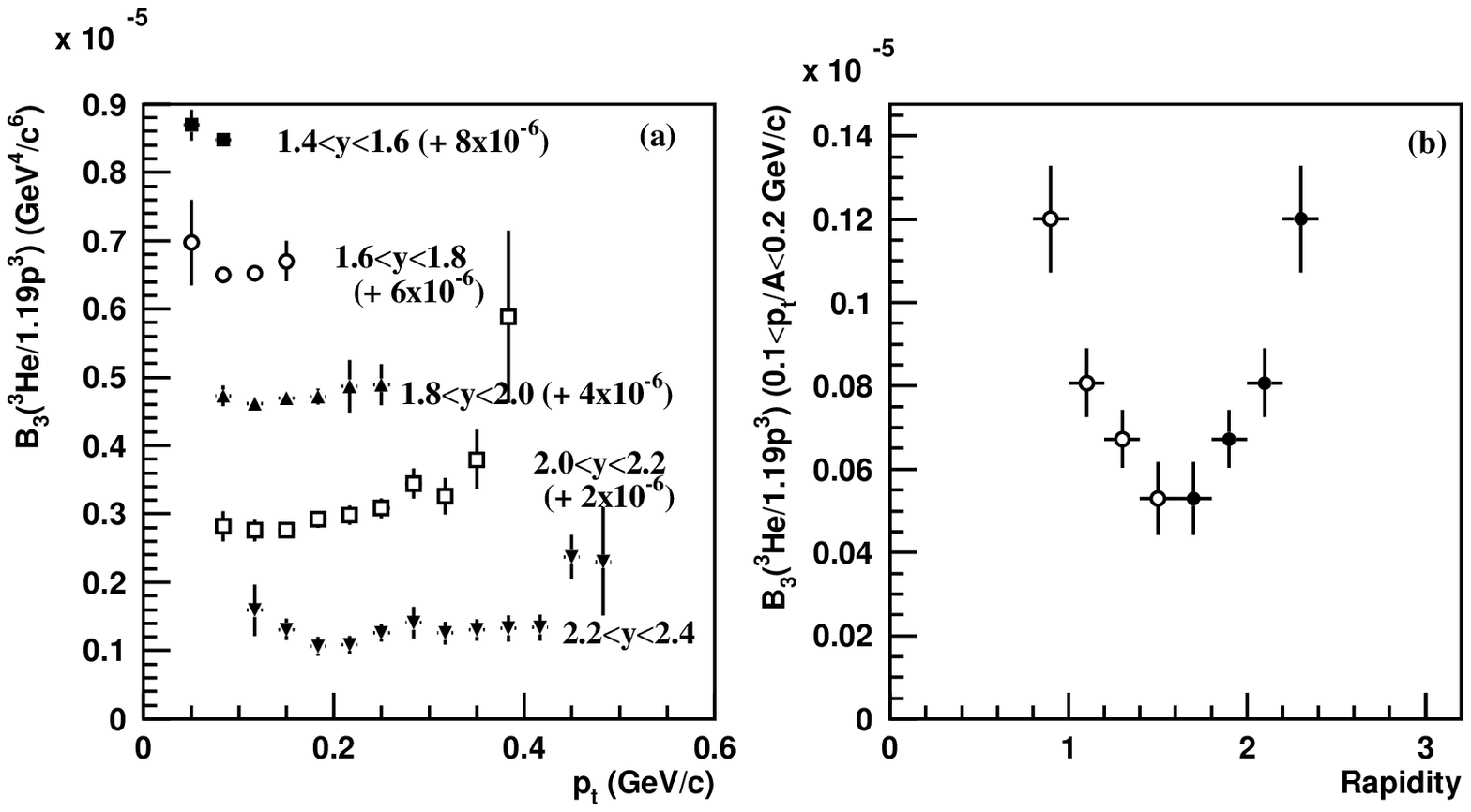}
\caption{
Coalescence $B_{3}$ parameters, with an assumed neutron to
proton ratio of 1.19, shown (a) 
as a function of transverse momentum 
in several rapidity slices and (b)
as a function of rapidity near $p_{T}=0$.  
In panel (a), error bars include only point-to-point errors
(a rapidity dependent error of 8\% and a global 11\% error are not
included).  In panel (b), hollow points represent reflections 
about center-of-mass rapidity.  Here, the rapidity-dependent errors 
are included but the 11\% global systematic error is not.
}
\label{fig:evb3}
\end{figure}

In Figures~\ref{fig:b2evan} and~\ref{fig:evb3} we show
measurements of $B_{2}$ and $B_{3}$ as a function of rapidity and transverse
momentum.  Because nucleons which are feed-down from 
weak decays are not available
as nucleons for coalescence, we have subtracted the contribution to proton
and neutron invariant yields from hyperon feed down for the 
calculation of $B_{A}$. 

The values of the $B_{A}$ in Figures~\ref{fig:b2evan} and~\ref{fig:evb3}
are clearly not momentum independent as is assumed in many early coalescence
models (\cite{sato} and \cite{bond}, for example).  This is expected
given the large amount of evidence for collective flow in heavy-ion collisions
\cite{pbm}.  The values of the $B_{A}$ parameters seem to increase
slightly with increasing transverse momentum and there is a clear increase
away from center-of-mass rapidity.  Both of these increases in $B_{A}$ away
from center-of-mass momentum are consistent with 
expectations from an expanding source, although the longitudinal
dependence can also be interpreted as a sign of incomplete
stopping.  Indeed the fact that the invariant yields of
antiprotons near $p_{T}$=0 are strongly peaked near center-of-mass
rapidity as measured by E864 \cite{pbarprc} while the proton yields 
are essentially flat may be taken as further evidence of incomplete stopping. 

\subsection{Source Size Calculations}
\label{sec-sizes}

The $B_{A}$ parameters can be related through coalescence models to source
sizes.  Following the model of Sato and Yazaki \cite{sato} 
which uses a density matrix
representation of the source distribution and projects it onto a representation
of the deuteron wave function, we can relate the $B_{A}$ parameters to  
the RMS source radius through
\begin{equation}
B_{A} = (\frac{2J_{A}+1}{2^{A}}) \frac{A^{5/2}}{m^{(A-1)}}(4 \pi \frac{\nu_{A} 
\nu}{\nu_{A} + \nu})^{(3/2)(A-1)}
\end{equation}
with $\nu_{A}$ the size parameter for a cluster with baryon number $A$ and
spin $J_{A}$ ($m$ represents the nucleon mass).
This model assumes the absence of collective motion of the nucleons, and
therefore a radius (given by $R_{RMS} = \sqrt{3/2 \nu}$) which is
independent of momentum.  We evaluate the source size using
$B_{2}$, $B_{3}$, $B_{4}$, $B_{6}$, and $B_{7}$ values from our measurements
nearest the collision center-of-mass (again assuming a neutron
to proton ratio of 1.19) and list the
results in Table~\ref{tab:satosiz}.  We have used values
for the $\nu_{A}$ from \cite{sato} for $A$ = 2,3 and 4 and
following \cite{llope} have done a polynomial extrapolation
to determine $\nu_{6}$ and $\nu_{7}$.   This extrapolation for
$\nu_{A}$ may be suspect particularly for the halo
nucleus $^{6}He$, but the final value for $R_{RMS}$ is quite
insensitive to the value for $\nu_{6}$.  The extracted radius for
deuterons $A$=2 is considerably larger than the initial size
of the colliding nuclei, and the radius parameters decrease
for clusters of increasing mass.

\begin{figure}
\centering\leavevmode\epsfbox[29 148 537 657]{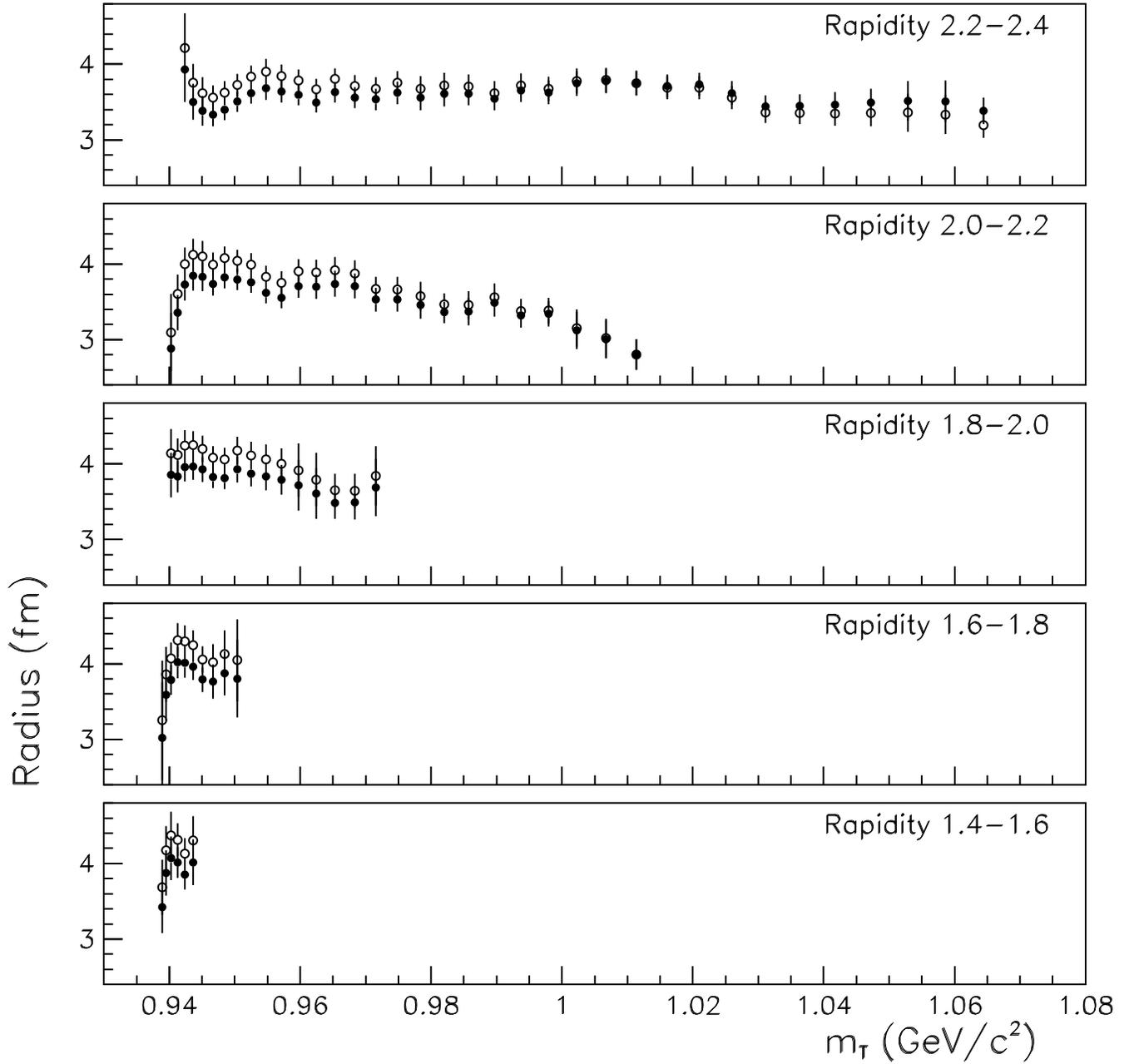}
\caption{Radius parameters extracted from measurements
of $B_{2}$ within the context of two different models.
The solid circles are to be understood as
$(R_{\perp}^{2}(m_{T})R_{\parallel}(m_{T}))^{1/3}$
in the context of the model of Scheibl and Heinz, while
the hollow circles are the extracted radii from the
cluster coalescence model of Llope et. al. }
\label{fig:radii}
\end{figure}

A model by Scheibl and Heinz \cite{heinzprc} which includes the 
effect of collective flow in a density matrix prescription for
coalescence leads to an expression for source dimensions:
\begin{equation}
B_{2} = \frac{3 \pi^{3/2} <C_{d}>} {2m_{T}R_{\perp}^{2}(m_{T})
R_{\parallel}(m_{T})} e^{2(m_{T}-m)(\frac{1}{T_{p}^{*}}-\frac{1}{T_{d}^{*}})}
\label{eq:heinz}
\end{equation}
where $R_{\perp}$ and $R_{\parallel}$ are the transverse and longitudinal
dimensions of the fraction of the source which contributes to 
deuteron emission (comparable to the radius parameters extracted in the
YKP parameterization of HBT interferometry), $<C_{d}>$ 
is a quantum mechanical correction factor
for the finite size of the deuteron which is evaluated by the authors under
various assumptions about the source, and $T_{p}^{*}$ and $T_{d}^{*}$ are
the inverse slope parameters for protons and deuterons.  
Equation~\ref{eq:heinz} as written assumes a box density 
profile for the source; a 
Gaussian profile would result in the absence of the final exponential 
factor.  Plugging in our results for $B_{2}$, we can extract values
for $(R_{\perp}^{2}(m_{T})R_{\parallel}(m_{T}))^{1/3}$ which are shown
in Figure~\ref{fig:radii} as the solid circles.  For the calculations
shown here we have used
0.75 for $<C_{d}>$ and the values for $T_{d}^{*}$ and $T_{p}^{*}$
as measured at rapidity 2.3.

Shown also in Figure~\ref{fig:radii} as hollow circles are the
results of source size calculations with the similar fragment coalescence
model of Llope {\it et. al.} \cite{llope} via the equation
\begin{equation}
R_{c}^{3} = \pi^{3/2} \frac{(2J_{c}+1)}{(2J_{a}+1)(2J_{b}+1)}
\frac{m_{c}}{m_{a}m_{b}}
\frac{E_{a}\frac{d^{3}N_{a}}{dp_{a}^{3}}E_{b}\frac{d^{3}N_{b}}
{dp_{b}^{3}}}{E_{c}\frac{d^{3}N_{c}}{dp_{c}^{3}}}
\label{eq:llope}
\end{equation}
which relates the effective source radius $R$ in the frame of
a cluster $c$ which may be formed through the coalescence of smaller 
clusters $a$ and $b$.
(Note that the radius parameters $R$ shown in Figure~\ref{fig:radii}
are meant to describe sources of the form $\rho(r) \propto 
exp(-r^{2}/2R^{2})$ and so correspond to RMS radii of 
$R_{RMS} = \sqrt{3}R$.)


We observe in Figure~\ref{fig:radii} that there is a 
decrease in source size with
increasing distance away from the center-of-mass (again,
as expected for a source with radial expansion), but
our measurements do not give a clear picture
concerning scaling of
the source size with transverse mass 
\cite{nkg_parkcity} such as has been
noted in results for sizes from HBT two-particle 
correlations and in measurements of $B_{2}$
at the CERN SPS \cite{mm}.  

\subsection{Comparison with RQMD}

We can also compare our results with predictions from the cascade
model RQMD \cite{RQMD} version 2.3.  The complex 
many-body processes by which light nuclei are formed are not included in 
RQMD, rather an afterburner
\cite{mati} is used to calculate the coalescence of these
states based upon an input of the positions and momenta of the
nucleons at freeze-out.  This model then explicitly includes
the position-momentum correlations due to expansion, etc. that
are present in RQMD.

\begin{figure}
\centering\leavevmode\epsfbox[51 137 520 672]{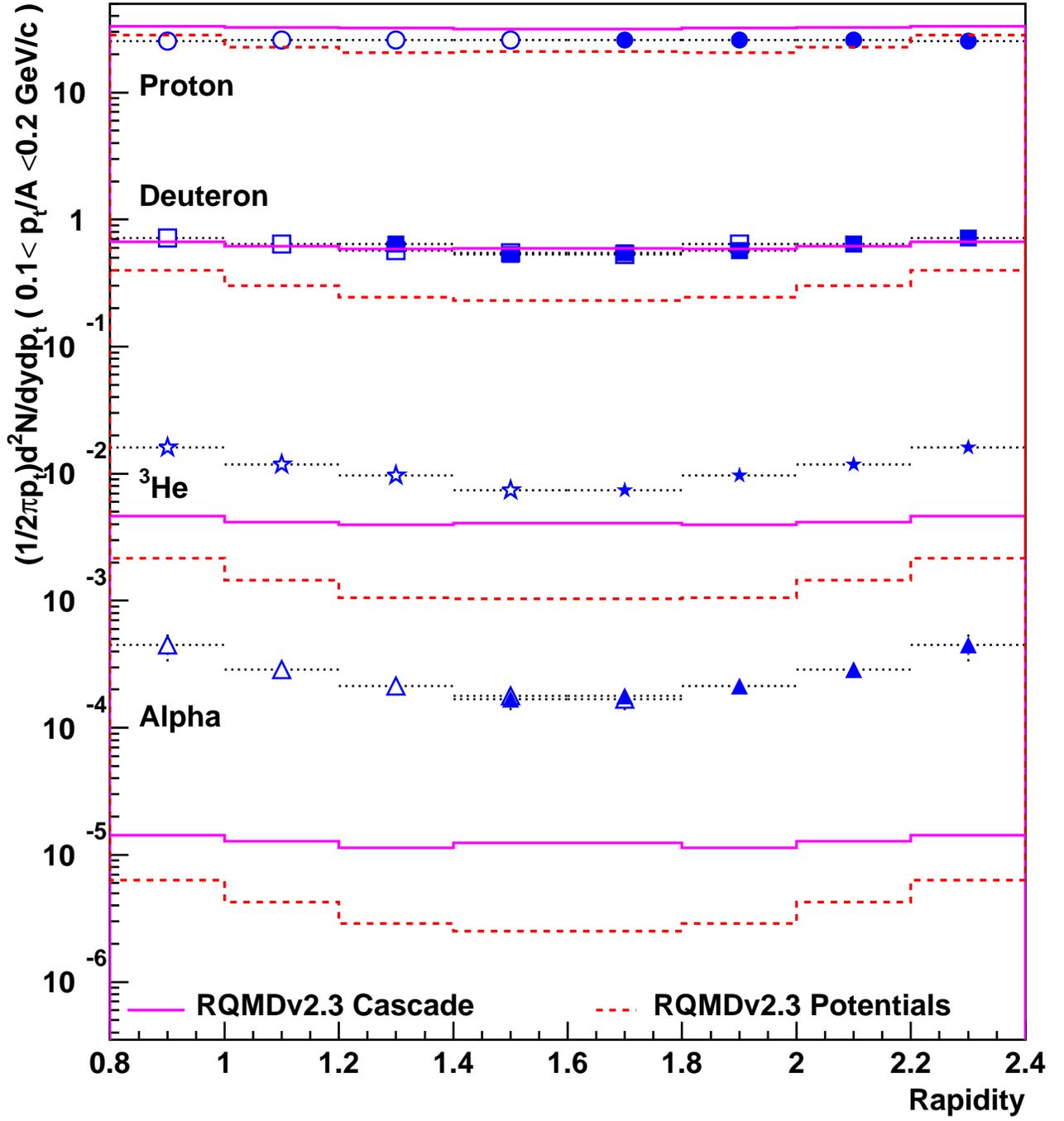}
\caption{Yields measured by E864 near $p_{T}=0$
(in units of $c^{2}/GeV^{2}$) as a function of rapidity for protons
(circles), deuterons (squares), $^{3}He$ nuclei (stars) and alpha 
particles (triangles).  For
comparison, the predictions of the model RQMDv2.3 with a 
coalescence afterburner are shown for the four species both for 
RQMD in cascade mode
(solid histogram lines) and with mean-field potentials (dashed 
histogram lines).  Hollow symbols represent reflections
of E864 data points about midrapidity.}
\label{fig:rqmd_rap}
\end{figure}

Figure~\ref{fig:rqmd_rap} displays the E864 measurements along with 
predictions from
RQMD with the coalescence afterburner 
for comparison over the transverse momentum range 
$0.1 \leq\ p_{T}/A \leq 0.2$ GeV/c as a function of rapidity.
For comparison, RQMDv2.3 was run under two different conditions,
one including the effect of repulsive mean-field potentials (potentials
mode) and one not (cascade mode).
We note that there is an increasing disagreement with increasing mass
as noted previously in Reference \cite{bennet}.  Calculations of
$^{4}He$ production in cascade mode are at a level of approximately 100 
lower than our measurements; with potentials on, the discrepancy 
is still larger by about a factor of two.  The level of disagreement
varies somewhat in with transverse momentum (the slopes are in fact
generally better predicted with potentials on than off) but 
particularly for the heavier states the change is slight 
compared to the overall level of disagreement.     

\subsection{Scaling of Yields Versus Mass}

\begin{figure}
\centering\leavevmode\epsfbox[12 165 531 655]{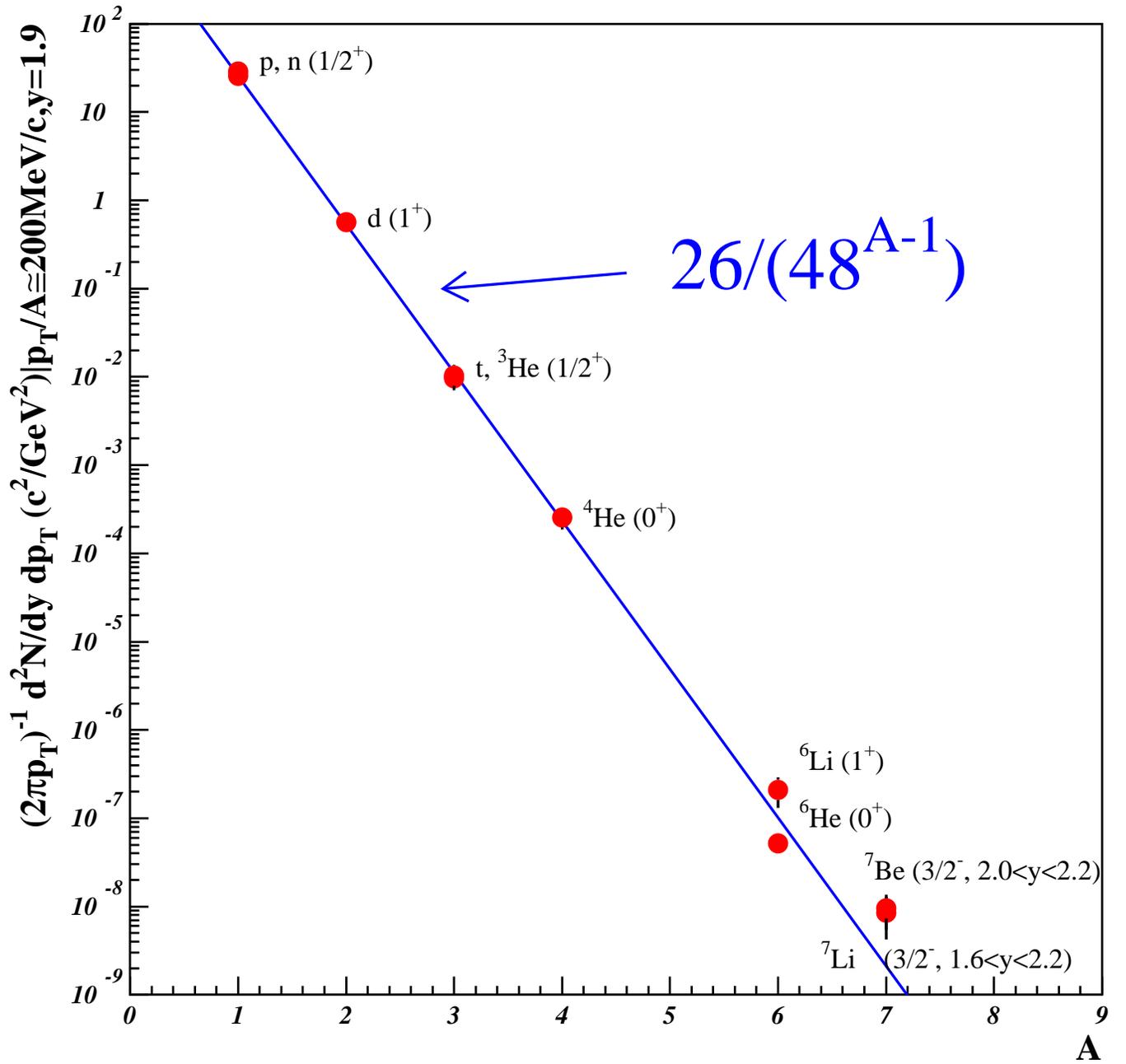}
\caption{Mass dependence of invariant yields of light nuclei from A=1 up
to A=7.  Yields are measured at $y$=1.9, $p_{T}/A \leq $ 300 MeV/c except where
otherwise noted in the plot.}
\label{fig:adep}
\end{figure}

Shown in Figure~\ref{fig:adep} are the invariant yields in a small kinematic
region at or near $y=1.9, p_{T} \leq 300 MeV$.  Over ten orders of magnitude,
the yields in this kinematic bin fit very closely to an exponential dependence
with a penalty factor of approximately 48 for each nucleon added (see 
\cite{eos} and references therein for a discussion of such exponential 
behavior of cluster yields
at lower collision energies).  Of course,
as we have seen above, the yields for different species have different kinematic
dependences due to collective motion, and so this should not be taken as
a penalty factor which governs the integrated yields of the 
various species, which from the kinematic dependences of the $B_{A}$ 
parameters discussed in Section~\ref{sec-ba} can clearly be different.

We can make a crude estimate of the difference in these two penalty
factors by parameterizing the rapidity and mass dependences of the 
inverse slope parameters as follows:  from our neutron measurements
\cite{neut} which extend up to beam rapidity, we can roughly 
parameterize the distribution of T in rapidity as a Gaussian with a width of 1.1
units.  From Figure~\ref{fig:tvsm}, we also can 
make a rough parameterization of the 
mass dependence of the inverse slope as  $T \propto (A+1.0)$
With these two parameterizations and our penalty factor of 48 at
$y=1.9, p_{T} \leq 300 MeV$, we calculate a penalty factor
of 25 in the integrated yields for the addition of an extra
baryon to a coalesced state.  This should be considered as
a lower limit on such a penalty factor since as previously noted
the inverse slope parameter rolls over as a function of mass rather than 
following the linear dependence we have used in 
this estimate \cite{xzb_ismd}.   

We can also estimate this penalty factor in overall yields by
following section VI of Reference \cite{heinzprc}.  Equation
6.10 in this reference allows us effectively to make an estimate
of the difference between the penalty factor near $p_{T}$ = 0 and
the overall penalty factor in two limiting cases: a static, homogeneous
fireball (which would translate our penalty factor of 48 at $p_{T}$
=0 to an overall penalty factor of 72) and a rapidly expanding
system (which would result in an overall penalty factor of 39).

\subsection{Spin and Isospin Dependences of Yields}

\begin{figure}
\centering\leavevmode\epsfbox[102 218 513 617]{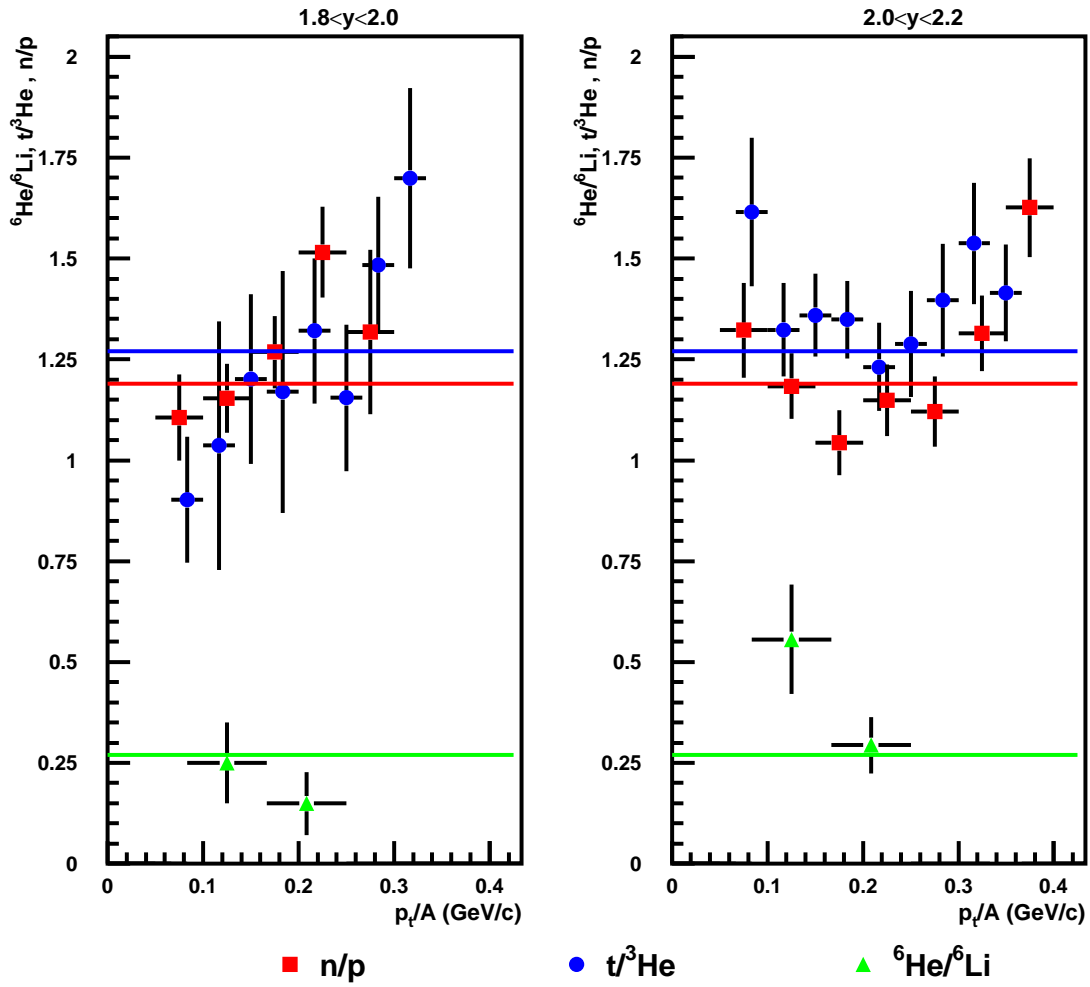}
\caption{$n/p$, $t/^{3}He$, and $^{6}He/^{6}Li$ ratios as a function of
transverse momentum in two different rapidity slices.}
\label{fig:spin}
\end{figure}

In Figure ~\ref{fig:spin}, we display three ratios as a function of transverse 
momentum in the rapidity ranges y=1.8-2.0 and y=2.0-2.2.  The
three ratios are the ratios of invariant yield of neutrons over
invariant yield of protons, the ratio of yield of tritons over
yield of $^{3}He$ and the ratio of $^{6}He$ to $^{6}Li$.  The
$n/p$ and $t/^{3}He$ ratios are consistent with a value
of approximately 1.2 (in Reference \cite{neut}, we extract values
of $1.19 \pm .08$ and $1.23 \pm .04$, respectively, 
for the two ratios in the range $1.6 \leq y \leq 2.4$.).  
In contrast, the $^{6}He/^{6}Li$ is much nearer to a value of 0.3.
$p$, $n$, $^{3}He$ and $t$ are all spin J=1/2 states while 
$^{6}He$ is spin 0 and $^{6}Li$ is spin 1.  We take this as evidence
that the yields scale as the degeneracy factor 2J+1 which is
commonly predicted in thermal and coalescence models.

With the dependences upon mass number, isospin and spin divided away
one can examine the yields for other dependences.  This topic, including
a possible dependence on binding energy per nucleon with an inverse slope
parameter dependence of a few MeV, is discussed in Reference \cite{xzbprl}.

\section{Summary}

We have shown results of measurements of light nuclei from $A$=1 
to $A$=7.  The increase with mass of light nuclei inverse slope
parameters appears to roll over at approximately $A$=3 in central
events but not necessarily in more peripheral events.  Also we
have shown that the yields near $p_{T}$=0 are concave as a 
function of rapidity and that
this relative concavity increases in more peripheral events and
in higher mass nuclei, consistent with both radial expansion and incomplete
stopping.  These trends are also evident from our observations
of the kinematic dependences of the $B_{A}$ parameters.  From these
parameters we have extracted source dimensions from various models.
Efforts to extract more quantitative information about the source
from these measurements using the cascade model RQMD with a 
coalescence afterburner were unsuccessful as predictions of
the model differ from our results by an amount that increases with
mass and reaches a level of 100 or more by $A$=4.

We have also examined the overall scaling of the yields up to $A$=7,
extracting a penalty factor of about 48 to add a nucleon to a
coalesced state near midrapidity at low transverse momentum. 
This likely translates into a somewhat smaller penalty factor
in overall yields for the addition of a nucleon, but we have
argued that this is unlikely to differ by as much as a factor
of two from our measured penalty of 48. 

\section{Acknowledgements}

We gratefully acknowledge the efforts of the AGS staff in providing the beam.  
This work was supported in part by grants from the Department of Energy 
(DOE) High Energy Physics Division, the DOE Nuclear Physics Division, and 
the National Science Foundation.

\section{APPENDIX}
Shown in Table~\ref{tab:mass1} through Table~\ref{tab:mass7} are results of
measurements by E864 of invariant multiplicities of light nuclei from $A$=1
through $A$=7 in 10\% most central Au+Pb collisions.  Results for less central
data are also listed for protons, deuterons and $^{3}He$.


\newpage


\newpage

\newpage

\newpage

\begin{figure}
\centering\leavevmode\epsfbox[23 147 529 658]{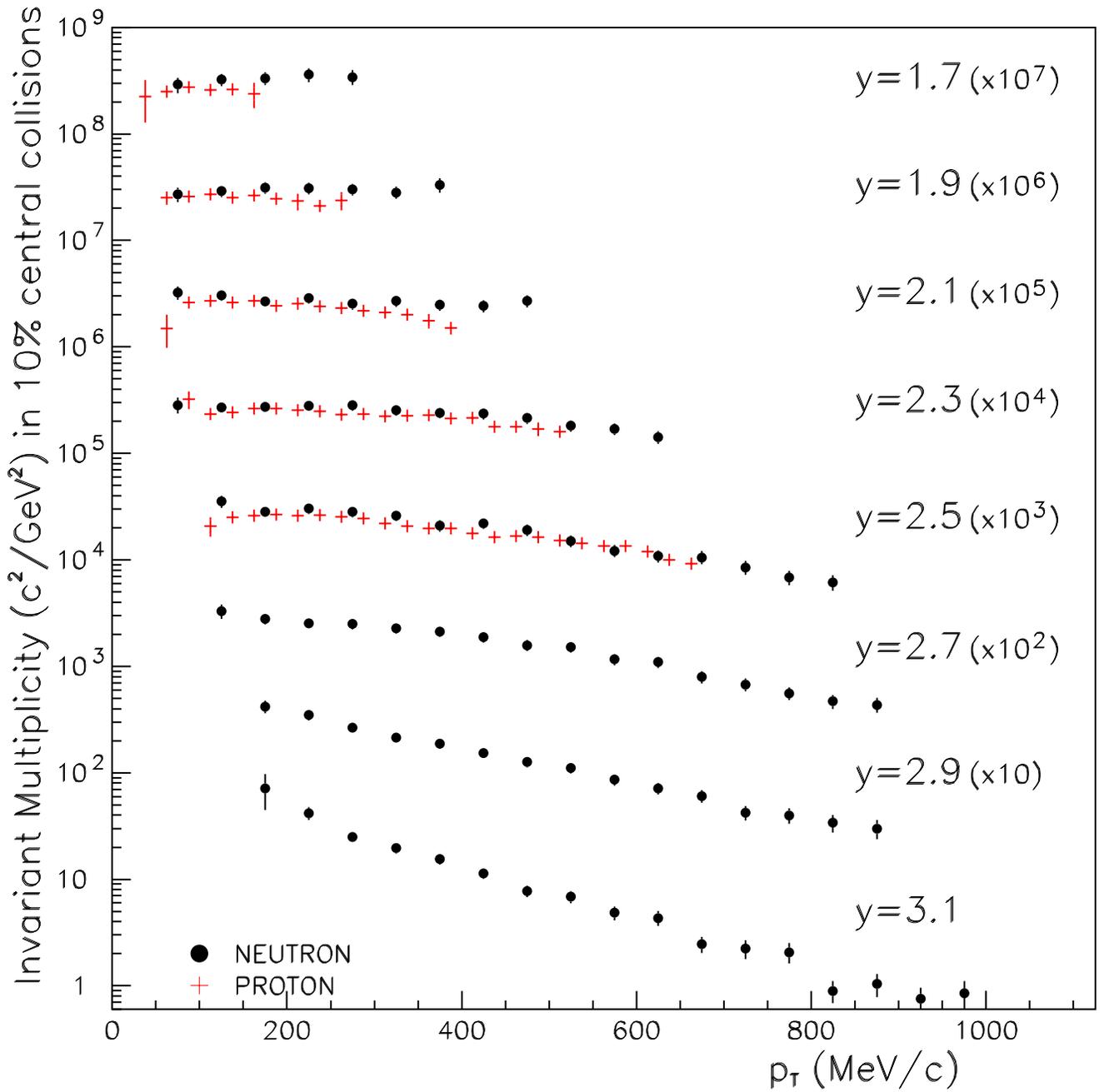}
\caption{Invariant yields of protons and neutrons in 10\% most
central Au+Pb collisions.}
\label{fig:inv_mult_np_2}
\end{figure}

\begin{figure}
\centering\leavevmode\epsfbox[69 181 513 653]{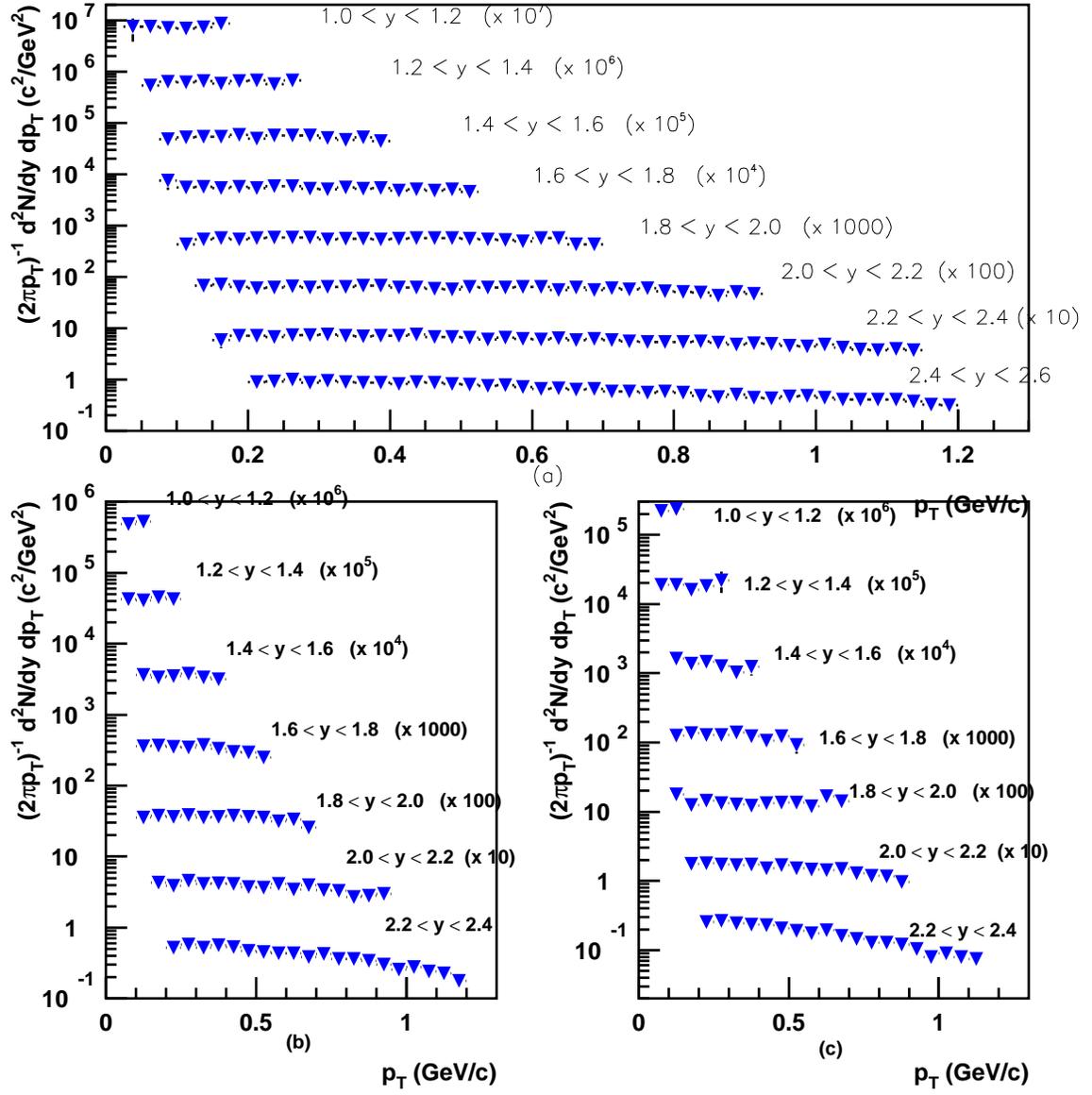}
\caption{Panel (a) displays invariant yields for deuterons in 10\% most
central Au+Pb collisions measured by E864.  Panels (b) and (c) 
display E864 measurements of deuteron invariant yields
for 10-38\% and 38-66\% central collisions, respectively. }
\label{fig:deut_mult}
\end{figure}

\begin{figure}
\centering\leavevmode\epsfbox[77 153 516 680]{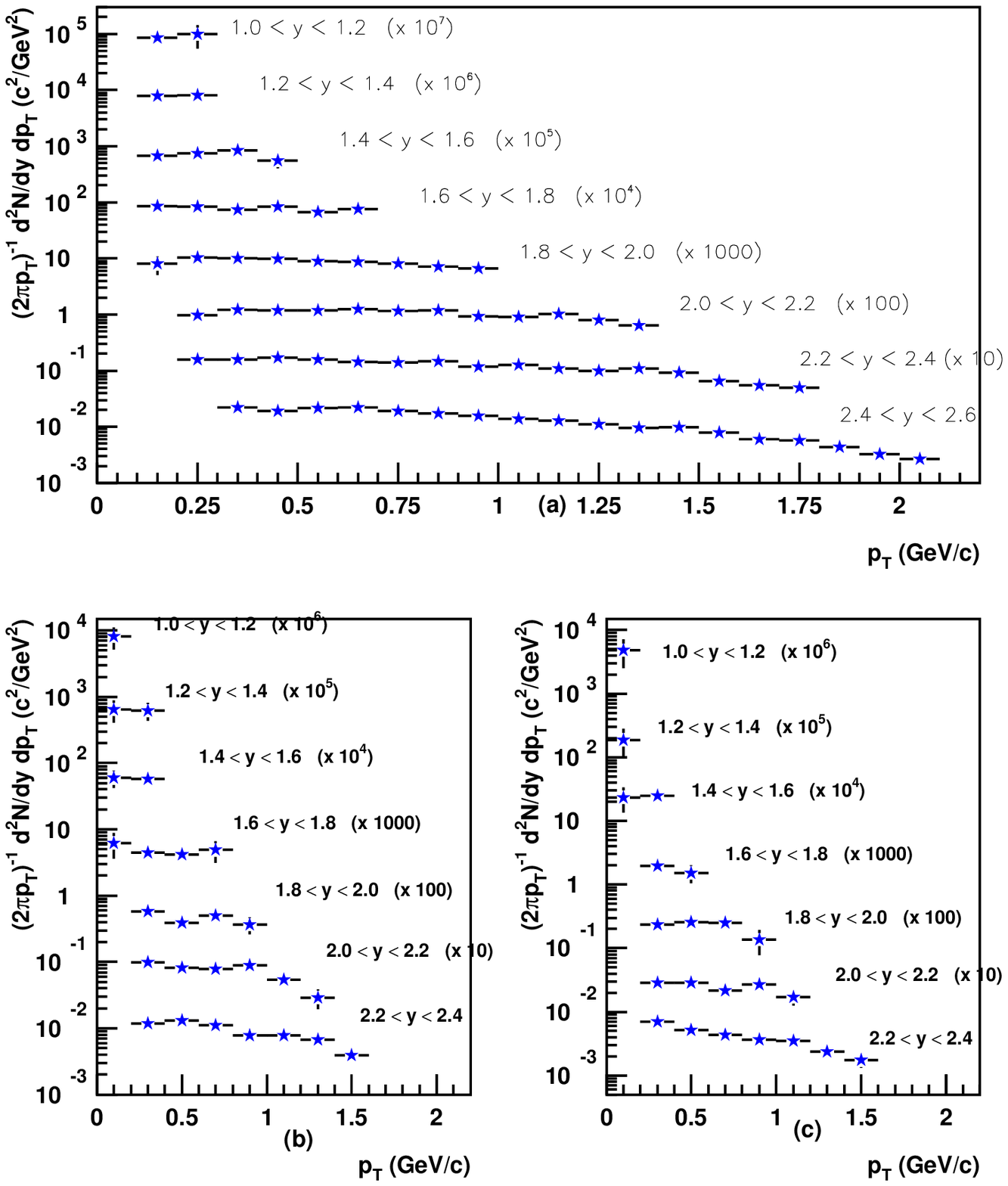}
\caption{Panel (a) displays invariant yields for $^{3}He$ nuclei
in 10\% most
central Au+Pb collisions.  Panels (b) and (c) display the same
for 10-38\% and 38-66\% central collisions, respectively.  }
\label{fig:he3_mult}
\end{figure}

\begin{figure}
\centering\leavevmode\epsfbox[87 412 508 636]{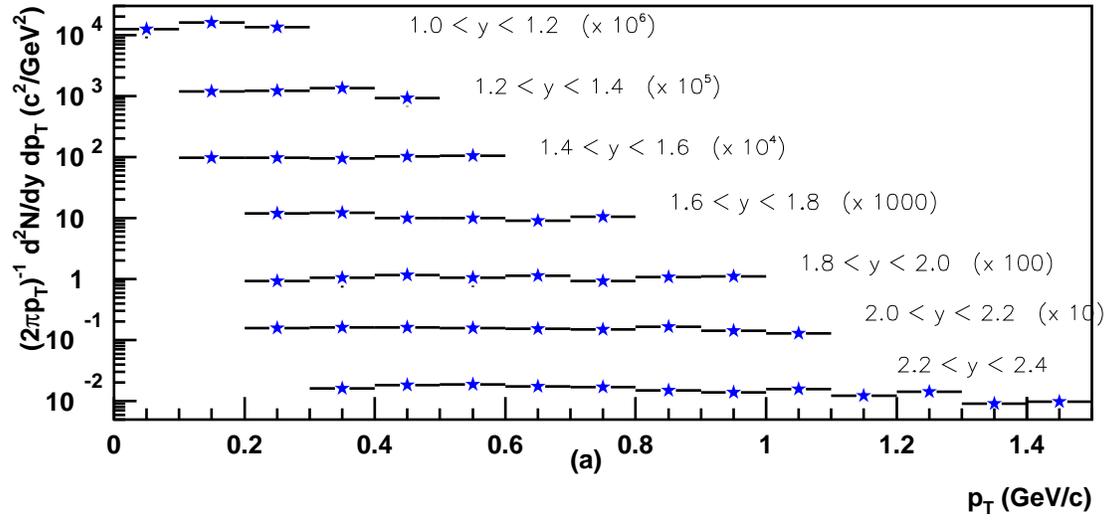}
\caption{Invariant yields for tritons in 10\% most
central Au+Pb collisions.}
\label{fig:trit_mult}
\end{figure}

\begin{figure}
\centering\leavevmode\epsfbox[20 154 527 649]{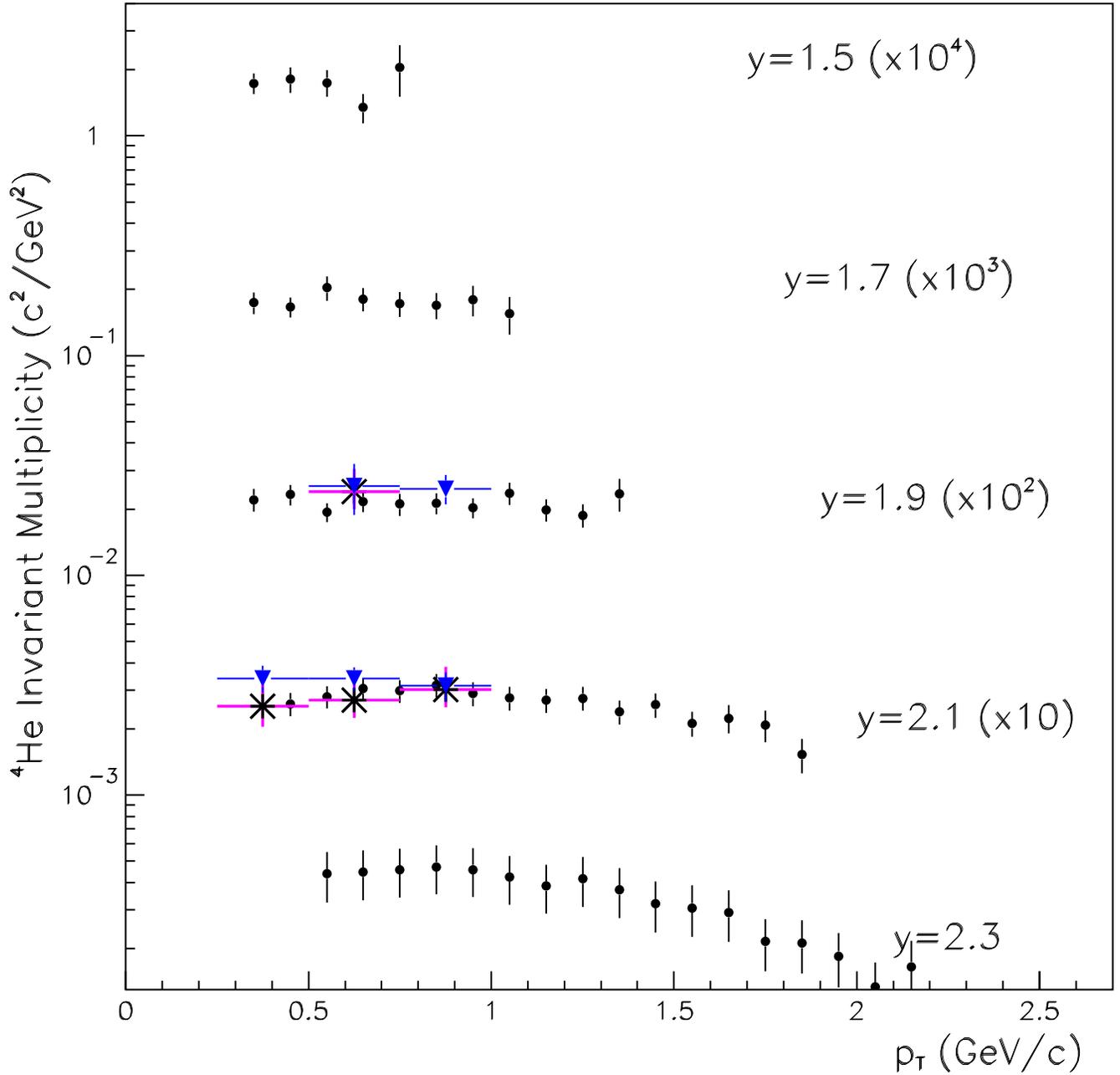}
\caption{Invariant yields for alpha particles in 10\% most
central collisions.  Solid circles represent measurements
from data taken in the 1998 run (with a +.45 T field in M1 and
M2) , triangles are measurements
from the 1996 run (+1.5T field) and stars are measurements from the
1995 run (+1.5 T field).  The larger uncertainties in the data points
at rapidity 2.3 are due to increased contamination from $^{3}He$ at this
rapidity. 1995 data is from Au+Pb while 1996 and 1998 data
are from Au+Pt.}
\label{fig:alpha_mult}
\end{figure}

\newpage

\newpage

%
%
\newpage

\begin{table}
\caption{Summary of experimental conditions under which the 
light nuclei measurements were made.  Events refers to
the number (in millions) of sampled events in a given data
set.}
\label{tab:datasum}
\begin{tabular}{|c|c|c|c|c|c|c|c|}
Year & Field (T) & Trigger  & Events (M) & Species               
                        
& Rapidity               & Centrality              \\ \tableline \tableline
1994 & +.75      & MULT     & 24         & p,d,$^{3}He$,$^{4}He$ 
                        
& $1.2 \leq y \leq 2.0 $ &  0-10\%                 \\ \tableline
1995 & +.45      & MULT     &  6         & p,d,$^{3}He$            
                      
& $1.2 \leq y \leq 2.4 $ &  0-10,10-38,38-66\%      \\ \tableline
1996 & +1.5      & MULT     &  7         & n                        
                     
& $1.6 \leq y \leq 3.2 $ &  0-10.10-38,38-66\%      \\ \tableline
1996 & +1.5      & MULT+LET & 13000      & $^{4}He$,$^{6}He$,$^{6}Li$,
$^{7}Li$,$^{7}Be$  
& $1.6 \leq y \leq 2.2 $ &  10\%                    \\ \tableline
1998 & +.45      & MULT+LET & 2000       & $^{4}He$                   
                   
& $1.4 \leq y \leq 2.4 $ &  10\%                    \\ 
\end{tabular}
\end{table}

\begin{table}
\caption{$B_{A}$ parameters in units of ($GeV^{2}/c^{3})^{(A-1)}$ 
near collision center-of-mass 
and source sizes extracted from them from the
model of Sato and Yazaki.}
\label{tab:satosiz}
\begin{tabular}{|c|c|c|c|c|}
A   &species &  $B_{A}$                      & $R_{RMS}(fm)$ 
& $y,p_{T}$                                          \\ \tableline \tableline
2   &   d    & $8.7 \pm 1.4 \times 10^{-5} $ & 14.8$\pm$ .7  
& $1.4 \leq y \leq 1.8,\frac{p_{T}}{A} \leq 200 MeV$ \\ \tableline
3   &$^{3}He$& $5.1 \pm 1.1 \times 10^{-7} $ & 12.2$\pm$ .5  
& $1.4 \leq y \leq 1.8,\frac{p_{T}}{A} \leq 200 MeV$ \\ \tableline
4   &$^{4}He$& $4.9 \pm 1.3 \times 10^{-10}$ & 10.1$\pm$ .3  
& $1.4 \leq y \leq 1.8,\frac{p_{T}}{A} \leq 200 MeV$ \\ \tableline
6   &$^{6}He$& $1.9 \pm 0.8 \times 10^{-16}$ &  9.9$\pm$ .3  
& $1.6 \leq y \leq 1.8,\frac{p_{T}}{A} \leq 170 MeV$ \\ \tableline
7   &$^{7}Be$& $1.4 \pm 0.7 \times 10^{-18}$ &  9.2$\pm$ .3  
& $1.6 \leq y \leq 2.2,\frac{p_{T}}{A} \leq 250 MeV$ \\ 
\end{tabular}
\end{table}

\begin{table}
\caption{Invariant yields for protons in 10\% most central
Au+Pb collisions in units of $c^{2}/GeV^{2}$  Errors listed 
are systematic and statistical combined
in quadrature.}
\label{tab:mass1}
\begin{tabular}{|c|c|c|c|c|c|c|}
     Rapidity   &  1.4-1.6       &   1.6-1.8      &   1.8-2.0      
& 2.0-2.2        &
    2.2-2.4     &  2.4-2.6       \\ \tableline \tableline
  $p_{T}$(MeV/c)&                &                &                
& 
                &                \\ \tableline
  25-50         & 17.9$\pm$  3.5 & 22.4$\pm$  9.6 &                
&                &
                &                \\ \tableline
  50-75         & 28.0$\pm$  3.5 & 25.0$\pm$  2.2 & 25.2$\pm$  3.5 
& 14.9$\pm$  5.1 &
                &                \\ \tableline
  75-100        & 26.0$\pm$  2.2 & 27.5$\pm$  2.4 & 25.8$\pm$  2.2 
& 26.2$\pm$  2.5 &
 32.1$\pm$  6.1 &                \\ \tableline
  100-125       & 25.2$\pm$  3.7 & 25.9$\pm$  1.9 & 27.1$\pm$  2.0 
& 27.0$\pm$  2.5 &
 23.3$\pm$  2.4 & 20.8$\pm$  4.3 \\ \tableline
  125-150       &                & 26.2$\pm$  3.0 & 25.4$\pm$  1.7 
& 26.2$\pm$  1.9 &
 24.2$\pm$  1.9 & 25.2$\pm$  2.3 \\ \tableline
  150-175       &                & 23.8$\pm$  6.4 & 26.4$\pm$  2.0 
& 27.0$\pm$  1.9 &
 26.4$\pm$  1.9 & 26.1$\pm$  2.1 \\ \tableline
  175-200       &                &                & 24.6$\pm$  1.9 
& 24.3$\pm$  1.8 &
 26.3$\pm$  1.9 & 26.5$\pm$  2.1 \\ \tableline
  200-225       &                &                & 23.4$\pm$  4.1 
& 25.4$\pm$  1.9 &
 25.4$\pm$  1.8 & 26.0$\pm$  1.9 \\ \tableline
  225-250       &                &                & 21.1$\pm$  2.2 
& 24.1$\pm$  1.9 &
 24.9$\pm$  1.7 & 26.2$\pm$  1.8 \\ \tableline
  250-275       &                &                & 23.8$\pm$  4.6 
& 23.1$\pm$  1.7 &
 23.2$\pm$  1.7 & 25.5$\pm$  1.9 \\ \tableline
  275-300       &                &                &                
& 21.9$\pm$  1.6 &
 23.5$\pm$  1.9 & 24.5$\pm$  1.7 \\ \tableline
  300-325       &                &                &                
& 21.1$\pm$  2.0 &
 22.3$\pm$  1.6 & 22.0$\pm$  1.6 \\ \tableline
  325-350       &                &                &                
& 20.0$\pm$  1.6 &
 22.6$\pm$  1.6 & 20.7$\pm$  1.6 \\ \tableline
  350-375       &                &                &                
& 17.6$\pm$  2.7 &
 22.8$\pm$  1.7 & 19.7$\pm$  1.5 \\ \tableline
  375-400       &                &                &                
& 15.1$\pm$  2.0 &
 21.3$\pm$  1.6 & 19.7$\pm$  1.5 \\ \tableline
  400-425       &                &                &                
&                &
 21.6$\pm$  1.6 & 17.8$\pm$  1.5 \\ \tableline
  425-450       &                &                &                
&                &
 17.8$\pm$  1.3 & 16.3$\pm$  1.3 \\ \tableline
  450-475       &                &                &                
&                &
 17.8$\pm$  1.5 & 16.7$\pm$  1.4 \\ \tableline
  475-500       &                &                &                
&                &
 17.0$\pm$  2.5 & 16.3$\pm$  1.4 \\ \tableline
  500-525       &                &                &                
&                &
 15.9$\pm$  1.4 & 15.3$\pm$  1.2 \\ \tableline
  525-550       &               &                &                
&                &
                & 14.4$\pm$  1.2 \\ \tableline
  550-575       &               &                &                
&                &
                & 13.5$\pm$  1.1 \\ \tableline
  575-600       &               &                &                
&                &
                & 13.5$\pm$  1.1 \\ \tableline
  600-625       &               &                &                
&                &
                & 12.0$\pm$  1.0 \\ \tableline
  625-650       &                &                &                
&                &
                & 10.0$\pm$  0.8 \\ \tableline
  650-675       &                &                &                
&                &
                &  9.2$\pm$  0.9 \\
\end{tabular}
\end{table}

\begin{table}
\caption{Invariant yields for protons in 10 to 38\% 
most central
Au+Pb collisions in units of $c^{2}/GeV^{2}$  Errors listed 
are systematic and statistical combined
in quadrature.}
\label{tab:mass1_1038}
\begin{tabular}{|c|c|c|c|c|c|}
     Rapidity   &  1.4-1.6       &   1.6-1.8      &   1.8-2.0      
& 2.0-2.2       &
    2.2-2.4     \\ \tableline \tableline

  $p_{T}$(MeV/c)&                &                &                
&               &
                \\ \tableline

25-50           & 14.3$\pm$  2.9 &                &
                &                &
                \\ \tableline
50-75           & 17.9$\pm$  2.3 & 17.7$\pm$  1.6 &
 16.6$\pm$  2.3 & 19.1$\pm$  7.4 &
                \\ \tableline
75-100          & 16.5$\pm$  1.4 & 16.2$\pm$  1.4 &
 16.7$\pm$  1.5 & 17.7$\pm$  1.9 &
 20.0$\pm$  4.2 \\ \tableline
100-125         & 16.8$\pm$  2.5 & 16.8$\pm$  1.3 &
 17.3$\pm$  1.3 & 19.2$\pm$  1.8 &
 17.4$\pm$  1.8 \\ \tableline
125-150         &                & 16.6$\pm$  1.9 &
 16.8$\pm$  1.1 & 16.6$\pm$  1.2 &
 17.2$\pm$  1.4 \\ \tableline
150-175         &                & 14.7$\pm$  4.0 &
 17.2$\pm$  1.3 & 17.1$\pm$  1.2 &
 18.6$\pm$  1.4 \\ \tableline
175-200         &                &                &
 14.4$\pm$  1.1 & 16.3$\pm$  1.2 &
 18.2$\pm$  1.4 \\ \tableline
200-225         &                &                &
 14.9$\pm$  2.6 & 16.5$\pm$  1.2 &
 17.1$\pm$  1.2 \\ \tableline
225-250         &                &                &
 13.8$\pm$  1.5 & 15.1$\pm$  1.2 &
 17.4$\pm$  1.2 \\ \tableline
250-275         &                &                &
 13.2$\pm$  2.6 & 15.3$\pm$  1.2 &
 15.5$\pm$  1.1 \\ \tableline
275-300         &                &                &
                & 14.5$\pm$  1.1 &
 14.5$\pm$  1.2 \\ \tableline
300-325         &                &                &
                & 13.0$\pm$  1.2 &
 14.5$\pm$  1.1 \\ \tableline
325-350         &                &                &
                & 12.1$\pm$  1.0 &
 14.9$\pm$  1.1 \\ \tableline
350-375         &                &                &
                & 11.3$\pm$  1.8 &
 14.5$\pm$  1.1 \\ \tableline
375-400         &                &                &
                & 10.2$\pm$  1.4 &
 14.2$\pm$  1.1 \\ \tableline
400-425         &                &                &
                &                &
 13.8$\pm$  1.0 \\ \tableline
425-450         &                &                &
                &                &                
 12.1$\pm$  0.9 \\ \tableline
450-475         &                &                &
                &                &
 10.4$\pm$  0.9 \\ \tableline
475-500         &                &                &
                &                &
 11.1$\pm$  1.6 \\ \tableline
500-525         &                &                &
                &                &
 10.3$\pm$  0.9 \\ \tableline
\end{tabular}
\end{table}

\begin{table}
\caption{Invariant yields for protons in 38 to 66\% 
most central
Au+Pb collisions in units of $c^{2}/GeV^{2}$  Errors listed 
are systematic and statistical combined
in quadrature.}
\label{tab:mass1_3866}
\begin{tabular}{|c|c|c|c|c|c|}
     Rapidity   &  1.4-1.6       &   1.6-1.8      &   1.8-2.0      
& 2.0-2.2       &
    2.2-2.4     \\ \tableline \tableline

  $p_{T}$(MeV/c)&                &                &                
&               &
                \\ \tableline
25-50           &  4.4$\pm$  0.9 &                &
                &                &
                \\ \tableline
50-75           &  6.9$\pm$  0.9 &
  7.1$\pm$  0.7 &  7.5$\pm$  1.1 &  7.5$\pm$  2.9 &
                \\ \tableline
75-100          &  6.9$\pm$  0.7 &  7.2$\pm$
  0.7 &  7.4$\pm$  0.7 &  7.6$\pm$  0.8 &
  9.2$\pm$  2.1 \\ \tableline
100-125         &  6.6$\pm$  1.0 &  7.0$\pm$  0.6 &
  7.3$\pm$  0.6 &  8.1$\pm$  0.8 &
  8.0$\pm$  0.9 \\ \tableline
125-150         &                &  7.3$\pm$  0.8 &
  7.0$\pm$  0.5 &  7.4$\pm$  0.6 &
  8.3$\pm$  0.7 \\ \tableline
150-175         &                &  5.6$\pm$  1.5 &
  7.1$\pm$  0.6 &  7.6$\pm$  0.6 &
  8.9$\pm$  0.7 \\ \tableline
175-200         &                &                &
  5.7$\pm$  0.5 &  6.8$\pm$  0.5 &
  8.5$\pm$  0.7 \\ \tableline
200-225         &                &                &
  6.3$\pm$  1.1 &  7.1$\pm$  0.5 &
  8.1$\pm$  0.6 \\ \tableline
225-250         &                &                &
  5.1$\pm$  0.6 &  6.6$\pm$  0.6 &
  8.0$\pm$  0.6 \\ \tableline
250-275         &                &                &
  5.8$\pm$  1.2 &  6.5$\pm$  0.5 &
  7.1$\pm$  0.5 \\ \tableline
275-300         &                &                &
                &  6.6$\pm$  0.5 &
  7.0$\pm$  0.6 \\ \tableline
300-325         &                &                &
                &  5.7$\pm$  0.6 &
  6.8$\pm$  0.5 \\ \tableline
325-350         &                &                &
                &  5.3$\pm$  0.5 &
  7.1$\pm$  0.5 \\ \tableline
350-375         &                &                &
                &  4.7$\pm$  0.7 &
  6.5$\pm$  0.5 \\ \tableline
375-400         &                &                &
                &  4.3$\pm$  0.6 &
  6.3$\pm$  0.5 \\ \tableline
400-425         &                &                &
                &                &
  6.0$\pm$  0.5 \\ \tableline
425-450         &                &                &
                &                &
  5.2$\pm$  0.4 \\ \tableline
450-475         &                &                &
                &                &
  4.6$\pm$  0.4 \\ \tableline
475-500         &                &                &
                &                &
  4.8$\pm$  0.7 \\ \tableline
500-525         &                &                &
                &                &
  3.9$\pm$  0.4 \\
\end{tabular}
\end{table}

\begin{table}
\caption{Invariant yields for deuterons in 10\% most central
Au+Pb collisions in units of $10^{-2} c^{2}/GeV^{2}$ in bins with
low transverse momentum (see also the following table for the high
transverse momentum data).  Errors listed 
are systematic and statistical combined
in quadrature.  }
\label{tab:mass2}
\begin{tabular}{|c|c|c|c|c|c|c|c|c|}
     Rapidity  &  1.0-1.2       &   1.2-1.4      &   1.4-1.6      
& 1.6-1.8        &
    1.8-2.0     &  2.0-2.2       &   2.2-2.4      &   2.4-2.6     
\\ \tableline \tableline
$p_{T}(MeV/c)$&                &                &                & 
                &                &                &                \\ \tableline
         37.5 & 74.7$\pm$ 35.7 &                &                
&                &
                &                &                &                \\ \tableline
         62.5 & 76.4$\pm$  8.8 & 53.7$\pm$ 11.0 &                
&                &
                &                &                &                \\ \tableline
         87.5 & 71.2$\pm$  6.4 & 63.1$\pm$  5.8 & 48.2$\pm$  6.4 
& 75.9$\pm$ 26.7 &
                &                &                &                \\ \tableline
        112.5 & 69.3$\pm$  6.2 & 61.7$\pm$  5.3 & 52.6$\pm$  5.8 
& 56.1$\pm$  9.3 &
 43.0$\pm$  9.7 &                &                &                \\ \tableline
        137.5 & 72.5$\pm$  8.6 & 66.0$\pm$  5.9 & 53.9$\pm$  5.9 
& 56.6$\pm$  5.2 &
 54.4$\pm$  5.6 & 68.1$\pm$ 14.0 &                &                \\ \tableline
        162.5 & 86.6$\pm$ 16.6 & 59.3$\pm$  4.7 & 53.2$\pm$  4.9 
& 54.4$\pm$  5.4 &
 56.4$\pm$  5.2 & 70.4$\pm$  6.9 & 59.1$\pm$ 18.2 &                \\ \tableline
        187.5 &                & 65.1$\pm$  5.3 & 58.0$\pm$  4.6 
& 56.5$\pm$  5.0 &
 53.5$\pm$  4.4 & 64.5$\pm$  6.0 & 70.6$\pm$  8.6 &                \\ \tableline
        212.5 &                & 67.5$\pm$  7.2 & 50.1$\pm$  4.3 
& 54.5$\pm$  5.2 &
 56.6$\pm$  4.9 & 60.5$\pm$  4.7 & 71.0$\pm$  7.0 & 89.7$\pm$ 13.7 \\ \tableline
        237.5 &                & 56.7$\pm$  8.4 & 57.3$\pm$  5.3 
& 59.4$\pm$  4.6 &
 58.0$\pm$  4.5 & 61.6$\pm$  4.7 & 66.5$\pm$  6.2 & 91.0$\pm$  9.6 \\ \tableline
        262.5 &                & 67.3$\pm$ 14.3 & 57.2$\pm$  5.0 
& 59.4$\pm$  4.7 &
 58.6$\pm$  4.2 & 65.1$\pm$  4.9 & 72.9$\pm$  5.9 & 98.7$\pm$  8.2 \\ \tableline
        287.5 &                &                & 56.3$\pm$  5.0 
& 52.1$\pm$  5.2 &
 58.0$\pm$  4.2 & 61.1$\pm$  4.3 & 74.0$\pm$  5.5 & 86.3$\pm$  6.8 \\ \tableline
        312.5 &                &                & 50.8$\pm$  4.7 
& 51.0$\pm$  4.5 &
 54.5$\pm$  4.6 & 64.3$\pm$  4.3 & 75.2$\pm$  5.5 & 95.1$\pm$  8.4 \\ \tableline
        337.5 &                &                & 46.3$\pm$  5.7 
& 55.1$\pm$  4.5 &
 56.1$\pm$  4.8 & 64.6$\pm$  4.4 & 72.2$\pm$  5.1 & 89.8$\pm$  9.3 \\ \tableline
        362.5 &                &                & 52.7$\pm$  8.7 
& 52.0$\pm$  4.2 &
 54.8$\pm$  5.5 & 66.6$\pm$  4.4 & 68.4$\pm$  6.0 & 85.8$\pm$  9.1 \\ \tableline
        387.5 &                &                & 44.1$\pm$ 11.3 
& 53.7$\pm$  4.5 &
 54.2$\pm$  4.5 & 68.2$\pm$  4.9 & 70.3$\pm$  5.1 & 87.2$\pm$  8.8 \\ \tableline
        412.5 &                &                &                
& 47.8$\pm$  4.3 &
 55.1$\pm$  5.0 & 61.8$\pm$  4.7 & 71.3$\pm$  5.0 & 82.0$\pm$  7.8 \\ \tableline
        437.5 &                &                &                
& 51.3$\pm$  4.4 &
 56.2$\pm$  4.6 & 62.0$\pm$  5.4 & 75.9$\pm$  5.2 & 88.1$\pm$  9.1 \\ \tableline
        462.5 &                &                &                
& 48.2$\pm$  5.0 &
 56.5$\pm$  4.3 & 58.3$\pm$  5.1 & 66.7$\pm$  4.8 & 85.7$\pm$  8.6 \\ \tableline
        487.5 &                &                &                
& 50.8$\pm$  7.9 &
 57.2$\pm$  4.4 & 57.6$\pm$  4.5 & 68.2$\pm$  5.2 & 82.0$\pm$  8.8 \\ \tableline
        512.5 &                &                &                
& 46.1$\pm$  6.4 &
 55.0$\pm$  4.3 & 64.4$\pm$  5.3 & 65.2$\pm$  5.2 & 80.3$\pm$  7.1 \\ \tableline
        537.5 &                &                &                
&                &
 54.1$\pm$  4.3 & 61.1$\pm$  5.7 & 59.5$\pm$  4.7 & 75.0$\pm$  8.0 \\
\end{tabular}
\end{table}

\begin{table}
\caption{Invariant yields for deuterons in 10\% most central
Au+Pb collisions in units of $10^{-2} c^{2}/GeV^{2}$ in bins with
high transverse momentum (see also the previous table for the low transverse
momentum data).  Errors listed 
are systematic and statistical combined
in quadrature.  }
\label{tab:mass2b}
\begin{tabular}{|c|c|c|c|c|c|c|c|c|}
     Rapidity  &  1.0-1.2       &   1.2-1.4      &   1.4-1.6      
& 1.6-1.8        &
    1.8-2.0     &  2.0-2.2       &   2.2-2.4      &   2.4-2.6     
\\ \tableline \tableline
$p_{T}(MeV/c)$&                &                &                & 
                &                &                &                \\ \tableline
        512.5 &                &                &                
& 46.1$\pm$  6.4 &
 55.0$\pm$  4.3 & 64.4$\pm$  5.3 & 65.2$\pm$  5.2 & 80.3$\pm$  7.1 \\ \tableline
        537.5 &                &                &                
&                &
 54.1$\pm$  4.3 & 61.1$\pm$  5.7 & 59.5$\pm$  4.7 & 75.0$\pm$  8.0 \\ \tableline
        562.5 &                &                &                
&                &
 51.7$\pm$  4.3 & 61.2$\pm$  7.7 & 64.8$\pm$  5.6 & 77.1$\pm$  8.6 \\ \tableline
        587.5 &                &                &                
&                &
 50.2$\pm$  4.4 & 64.4$\pm$  4.9 & 62.1$\pm$  5.3 & 71.1$\pm$  7.4 \\ \tableline
        612.5 &                &                &                
&                &
 57.6$\pm$  4.9 & 63.5$\pm$  6.1 & 60.4$\pm$  5.2 & 65.9$\pm$  7.3 \\ \tableline
        637.5 &                &                &                
&                &
 56.3$\pm$  8.1 & 55.9$\pm$  4.2 & 63.6$\pm$  5.6 & 66.6$\pm$  8.4 \\ \tableline
        662.5 &                &                &                
&                &
 43.6$\pm$  4.4 & 61.8$\pm$  5.6 & 58.8$\pm$  5.2 & 63.7$\pm$  7.9 \\ \tableline
        687.5 &                &                &                
&                &
 43.0$\pm$ 10.0 & 55.7$\pm$  4.4 & 61.8$\pm$  5.8 & 65.7$\pm$  9.6 \\ \tableline
        712.5 &                &                &                
&                &
                & 59.5$\pm$  4.9 & 58.1$\pm$  5.1 & 58.4$\pm$  8.2 \\ \tableline
        737.5 &                &                &                
&                &
                & 57.4$\pm$  4.6 & 55.8$\pm$  4.6 & 58.0$\pm$  8.2 \\ \tableline
        762.5 &                &                &                
&                &
                & 60.2$\pm$  4.5 & 53.1$\pm$  4.5 & 55.4$\pm$  8.4 \\ \tableline
        787.5 &                &                &                
&                &
                & 52.6$\pm$  4.0 & 53.8$\pm$  4.2 & 58.6$\pm$  9.2 \\ \tableline
        812.5 &                &                &                
&                &
                & 50.7$\pm$  6.9 & 54.3$\pm$  4.3 & 55.8$\pm$  8.2 \\ \tableline
        837.5 &                &                &                
&                &
                & 49.3$\pm$  4.2 & 55.1$\pm$  4.7 & 47.6$\pm$  6.9 \\ \tableline
        862.5 &                &                &                
&                &
                & 43.3$\pm$  4.9 & 52.8$\pm$  4.4 & 45.5$\pm$  5.8 \\ \tableline
        887.5 &                &                &                
&                &
                & 51.3$\pm$  7.0 & 48.4$\pm$  4.1 & 50.4$\pm$  6.7 \\ \tableline
        912.5 &                &                &                
&                &
                & 47.0$\pm$  8.4 & 50.5$\pm$  4.4 & 44.4$\pm$  6.3 \\ \tableline
        937.5 &                &                &                
&                &
                &                & 49.4$\pm$  4.3 & 42.6$\pm$  5.3 \\ \tableline
        962.5 &                &                &                
&                &
                &                & 45.7$\pm$  3.8 & 46.2$\pm$  6.9 \\ \tableline
        987.5 &                &                &                
&                &
                &                & 44.1$\pm$  3.9 & 47.6$\pm$  7.5 \\ \tableline
       1012.5 &                &                &                
&                &
                &                & 48.2$\pm$  4.0 & 44.2$\pm$  6.9 \\ \tableline
       1037.5 &                &                &                
&                &
                &                & 42.2$\pm$  4.0 & 40.7$\pm$  6.6 \\ \tableline
       1062.5 &                &                &                
&                &
                &                & 38.1$\pm$  3.6 & 40.6$\pm$  6.5 \\ \tableline
       1087.5 &                &                &                
&                &
                &                & 37.0$\pm$  4.6 & 40.9$\pm$  5.3 \\ \tableline
       1112.5 &                &                &                
&                &
                &                & 39.9$\pm$  4.1 & 40.7$\pm$  5.9 \\ \tableline
       1137.5 &                &                &                
&                &
                &                & 37.6$\pm$  4.3 & 36.8$\pm$  4.8 \\ \tableline
       1162.5 &                &                &                
&                &
                &                &                & 32.0$\pm$  4.7 \\ \tableline
       1187.5 &                &                &                
&                &
                &                &                & 31.9$\pm$  5.1 \\ 
\end{tabular}
\end{table}

\begin{table}
\caption{Invariant yields for deuterons in 10\% to 38\%most central
Au+Pb collisions in units of $10^{-2} c^{2}/GeV^{2}$.  Errors listed 
are systematic and statistical combined
in quadrature.  }
\label{tab:mass2_1038}
\begin{tabular}{|c|c|c|c|c|c|c|c|}
     Rapidity   &  1.0-1.2       &   1.2-1.4      &   1.4-1.6      &
 1.6-1.8        &  1.8-2.0       &  2.0-2.2       &
   2.2-2.4      \\ \tableline \tableline
$p_{T}(MeV/c)$  &                &                &                &
                &                &                &
                \\ \tableline

         25     &                &                &                &
                &                &                &
                \\ \tableline
         75     & 48.2$\pm$  5.2 & 42.5$\pm$  5.5 &                &
                &                &                &
                \\ \tableline
        125     & 52.4$\pm$  5.3 & 40.6$\pm$  3.8 & 36.0$\pm$  4.2 &
 36.1$\pm$  4.7 & 35.2$\pm$  4.9 &                &
                \\ \tableline
        175     &                & 44.9$\pm$  4.5 & 33.6$\pm$  3.6 &
 36.9$\pm$  3.5 & 38.1$\pm$  3.5 & 42.7$\pm$  4.6 &
                \\ \tableline
        225     &                & 42.8$\pm$  5.4 & 34.6$\pm$  3.3 &
 35.2$\pm$  3.0 & 36.4$\pm$  3.2 & 38.5$\pm$  4.0 &
 52.3$\pm$  4.9 \\ \tableline
        275     &                &                & 38.2$\pm$  3.7 &
 34.4$\pm$  3.3 & 38.5$\pm$  2.9 & 45.3$\pm$  3.5 &
 57.7$\pm$  4.3 \\ \tableline
        325     &                &                & 33.4$\pm$  3.7 &
 37.7$\pm$  3.2 & 35.5$\pm$  3.6 & 41.3$\pm$  3.4 &
 52.1$\pm$  4.7 \\ \tableline
        375     &                &                & 31.4$\pm$  6.9 &
 33.0$\pm$  3.1 & 36.2$\pm$  3.4 & 42.3$\pm$  3.2 &
 56.4$\pm$  4.1 \\ \tableline
        425     &                &                &                &
 29.9$\pm$  2.9 & 37.8$\pm$  3.7 & 40.9$\pm$  3.4 &
 53.0$\pm$  3.9 \\ \tableline
        475     &                &                &                &
 29.4$\pm$  4.1 & 35.9$\pm$  3.1 & 37.0$\pm$  3.3 &
 46.8$\pm$  3.9 \\ \tableline
        525     &                &                &                &
 25.0$\pm$  5.1 & 35.7$\pm$  3.0 & 36.9$\pm$  4.6 &
 45.2$\pm$  4.1 \\ \tableline
        575     &                &                &                &
                & 31.5$\pm$  3.1 & 41.2$\pm$  3.5 &
 43.3$\pm$  4.2 \\ \tableline
        625     &                &                &                &
                & 33.6$\pm$  3.9 & 34.2$\pm$  4.0 &
 43.3$\pm$  3.8 \\ \tableline
        675     &                &                &                &
                & 25.8$\pm$  4.3 & 39.3$\pm$  3.3 &
 38.7$\pm$  3.5 \\ \tableline
        725     &                &                &                &
                &                & 33.6$\pm$  3.2 &
 42.2$\pm$  3.7 \\ \tableline
        775     &                &                &                &
                &                & 33.1$\pm$  3.6 &
 36.3$\pm$  3.1 \\ \tableline
        825     &                &                &                &
                &                & 27.1$\pm$  3.5 &
 36.1$\pm$  3.9 \\ \tableline
        875     &                &                &                &
                &                & 28.3$\pm$  3.6 &
 34.2$\pm$  3.5 \\ \tableline
        925     &                &                &                &
                &                & 29.6$\pm$  4.2 &
 30.3$\pm$  3.9 \\ \tableline
        975     &                &                &                &
                &                &                &
 25.7$\pm$  2.4 \\ \tableline
       1025     &                &                &                &
                &                &                &
 27.6$\pm$  2.5 \\ \tableline
       1075     &                &                &                &
                &                &                &
 24.0$\pm$  2.5 \\ \tableline
       1125     &                &                &                &
                &                &                &
 22.8$\pm$  3.1 \\ \tableline
       1175     &                &                &                &
                &                &                &
 17.6$\pm$  2.8 \\ 
\end{tabular}
\end{table}

\begin{table}
\caption{Invariant yields for deuterons in 38\% to 66\%most central
Au+Pb collisions in units of $10^{-2} c^{2}/GeV^{2}$.  Errors listed 
are systematic and statistical combined
in quadrature.  }
\label{tab:mass2_3866}
\begin{tabular}{|c|c|c|c|c|c|c|c|}
     Rapidity   &  1.0-1.2       &   1.2-1.4      &   1.4-1.6      &
 1.6-1.8        &  1.8-2.0       &  2.0-2.2       &
   2.2-2.4      \\ \tableline \tableline

$p_{T}(MeV/c)$  &                &                &                &
                &                &                &
                \\ \tableline
         75     & 22.3$\pm$  2.8 & 19.0$\pm$  2.8 &                &
                &                &                &
                \\ \tableline
        125     & 23.4$\pm$  2.8 & 18.9$\pm$  2.0 & 16.4$\pm$  2.1 &
 12.5$\pm$  1.9 & 17.7$\pm$  2.9 &                &
                \\ \tableline
        175     &                & 15.9$\pm$  2.0 & 13.8$\pm$  1.6 &
 13.5$\pm$  1.9 & 12.4$\pm$  1.7 & 17.6$\pm$  2.0 &
                \\ \tableline
        225     &                & 18.1$\pm$  2.9 & 14.8$\pm$  1.7 &
 13.0$\pm$  1.6 & 14.3$\pm$  2.1 & 18.1$\pm$  1.9 &
 25.4$\pm$  2.5 \\ \tableline
        275     &                & 22.0$\pm$  7.5 & 12.7$\pm$  1.7 &
 12.8$\pm$  1.7 & 13.2$\pm$  1.4 & 17.3$\pm$  1.6 &
 26.8$\pm$  2.3 \\ \tableline
        325     &                &                & 10.3$\pm$  1.6 &
 14.0$\pm$  1.8 & 12.9$\pm$  1.5 & 16.8$\pm$  1.5 &
 24.3$\pm$  2.1 \\ \tableline
        375     &                &                & 12.4$\pm$  3.2 &
 12.2$\pm$  1.7 & 12.2$\pm$  1.6 & 17.1$\pm$  1.6 &
 23.2$\pm$  2.0 \\ \tableline
        425     &                &                &                &
 10.6$\pm$  1.4 & 13.4$\pm$  1.5 & 15.2$\pm$  1.5 &
 22.7$\pm$  1.8 \\ \tableline
        475     &                &                &                &
 12.3$\pm$  2.1 & 13.6$\pm$  1.4 & 17.0$\pm$  1.6 &
 20.8$\pm$  1.8 \\ \tableline
        525     &                &                &                &
  9.2$\pm$  2.4 & 13.5$\pm$  1.4 & 15.2$\pm$  1.5 &
 19.1$\pm$  2.2 \\ \tableline
        575     &                &                &                &
                & 11.8$\pm$  1.3 & 14.7$\pm$  1.9 &
 17.6$\pm$  1.8 \\ \tableline
        625     &                &                &                &
                & 16.6$\pm$  2.2 & 14.4$\pm$  1.5 &
 19.6$\pm$  1.9 \\ \tableline
        675     &                &                &                &
                & 14.0$\pm$  2.6 & 15.1$\pm$  1.5 &
 16.2$\pm$  1.9 \\ \tableline
        725     &                &                &                &
                &                & 13.0$\pm$  1.4 &
 14.4$\pm$  2.4 \\ \tableline
        775     &                &                &                &
                &                & 12.0$\pm$  1.3 &
 12.8$\pm$  1.9 \\ \tableline
        825     &                &                &                &
                &                & 11.7$\pm$  1.5 &
 12.7$\pm$  1.4 \\ \tableline
        875     &                &                &                &
                &                &  9.7$\pm$  1.5 &
 12.1$\pm$  1.9 \\ \tableline
        925     &                &                &                &
                &                &                &
 10.4$\pm$  2.0 \\ \tableline
        975     &                &                &                &
                &                &                &
  7.9$\pm$  1.4 \\ \tableline
       1025     &                &                &                &
                &                &                &
  9.0$\pm$  1.5 \\ \tableline
       1075     &                &                &                &
                &                &                &
  7.9$\pm$  1.6 \\ \tableline
       1125     &                &                &                &
                &                &                &
  7.5$\pm$  1.1 \\ 
\end{tabular}
\end{table}

\begin{table}
\caption{Invariant yields for $^{3}He$ in 10\% most central
Au+Pb collisions in units of $10^{-3} c^{2}/GeV^{2}$.  Errors listed 
are systematic and statistical combined
in quadrature.  }
\label{tab:helium3}
\begin{tabular}{|c|c|c|c|c|c|c|c|c|}
     Rapidity  &  1.0-1.2       &   1.2-1.4      &   1.4-1.6      
& 1.6-1.8        &
    1.8-2.0     &  2.0-2.2       &   2.2-2.4      &   2.4-2.6     
\\ \tableline \tableline
$p_{T}(MeV/c)$&                &                &                & 
                &                &                &                \\ \tableline
         150 &  8.6$\pm$  1.6 &  7.9$\pm$  1.0 &  6.7$\pm$  1.0 
&  8.6$\pm$  1.7 &
  8.0$\pm$  2.9 &                &                &                \\ \tableline
         250 &  9.9$\pm$  4.4 &  8.1$\pm$  1.0 &  7.4$\pm$  0.7 
&  8.4$\pm$  0.8 &
 10.3$\pm$  1.1 &  9.8$\pm$  1.3 & 15.7$\pm$  3.8 &                \\ \tableline
         350 &                &                &  8.4$\pm$  1.0 
&  7.4$\pm$  0.7 &
 10.1$\pm$  0.8 & 12.1$\pm$  1.1 & 15.8$\pm$  1.5 & 22.2$\pm$  3.2 \\ \tableline
         450 &                &                &  5.6$\pm$  1.6 
&  8.3$\pm$  0.9 &
  9.7$\pm$  0.8 & 11.7$\pm$  1.0 & 16.8$\pm$  1.3 & 19.1$\pm$  1.8 \\ \tableline
         550 &                &                &                
&  6.6$\pm$  0.8 &
  8.9$\pm$  0.9 & 11.7$\pm$  0.9 & 15.6$\pm$  1.2 & 21.6$\pm$  1.8 \\ \tableline
         650 &                &                &                
&  7.6$\pm$  1.4 &
  8.6$\pm$  0.9 & 12.4$\pm$  1.0 & 14.3$\pm$  1.1 & 21.9$\pm$  1.8 \\ \tableline
         750 &                &                &                
&                &
  8.0$\pm$  0.9 & 11.5$\pm$  1.0 & 14.0$\pm$  1.1 & 19.2$\pm$  1.6 \\ \tableline
         850 &                &                &                
&                &
  7.2$\pm$  0.9 & 11.9$\pm$  1.2 & 14.7$\pm$  1.3 & 17.2$\pm$  1.6 \\ \tableline
         950 &                &                &               
 &                &
  6.6$\pm$  1.4 &  9.2$\pm$  0.9 & 11.8$\pm$  1.1 & 15.7$\pm$  1.7 \\ \tableline
        1050 &                &                &                
&                &
                &  9.1$\pm$  0.9 & 12.7$\pm$  1.1 & 13.9$\pm$  1.7 \\ \tableline
        1150 &                &                &                
&                &
                & 10.2$\pm$  1.1 & 10.9$\pm$  1.0 & 12.9$\pm$  1.6 \\ \tableline
        1250 &                &                &                
&                &
                &  8.1$\pm$  1.3 &  9.9$\pm$  0.9 & 11.2$\pm$  1.3 \\ \tableline
        1350 &                &                &                
&                &
                &  6.3$\pm$  1.2 & 10.8$\pm$  1.0 &  9.6$\pm$  1.1 \\ \tableline
        1450 &                &                &                
&                &
                &                &  9.3$\pm$  0.9 &  9.8$\pm$  1.0 \\ \tableline
        1550 &                &                &                
&                &
                &                &  6.5$\pm$  0.8 &  7.9$\pm$  0.9 \\ \tableline
        1650 &                &                &                
&                &
                &                &  5.4$\pm$  0.9 &  6.1$\pm$  0.8 \\ \tableline
        1750 &                &                &                
&                &
                &                &  5.0$\pm$  0.9 &  5.7$\pm$  0.7 \\ \tableline
        1850 &                &                &                
&                &
                &                &                &  4.4$\pm$  0.6 \\ \tableline
        1950 &                &                &                
&                &
                &                &                &  3.2$\pm$  0.5 \\ \tableline
        2050 &                &                &                
&                &
                &                &                &  2.6$\pm$  0.4 \\
\end{tabular}
\end{table}

\begin{table}
\caption{Invariant yields for $^{3}He$ in 10\% to 38\% most central
Au+Pb collisions in units of $10^{-3} c^{2}/GeV^{2}$.  Errors listed 
are systematic and statistical combined
in quadrature.  }
\label{tab:helium3_1038}
\begin{tabular}{|c|c|c|c|c|c|c|c|}
     Rapidity   &    1.0-1.2     &   1.2-1.4      &   1.4-1.6      &
  1.6-1.8       &    1.8-2.0     &   2.0-2.2      &   2.2-2.4     
\\ \tableline \tableline
$p_{T}(MeV/c)$  &                &                &                & 
                &                &                &                
\\ \tableline
         100    &  8.2$\pm$  3.0 &  6.4$\pm$  2.3 &  6.0$\pm$  1.8 &
  6.2$\pm$  2.6 &                &                &                
\\ \tableline
         300    &                &  6.2$\pm$  1.8 &  5.8$\pm$  1.0 &
  4.5$\pm$  0.7 &  5.8$\pm$  0.9 &  9.8$\pm$  1.3 & 11.9$\pm$  2.0 
\\ \tableline
         500    &                &                &                &
  4.2$\pm$  0.9 &  3.9$\pm$  0.6 &  8.2$\pm$  0.9 & 13.2$\pm$  1.2 
\\ \tableline
         700    &                &                &                &
  4.9$\pm$  1.8 &  5.0$\pm$  1.0 &  7.8$\pm$  1.0 & 11.1$\pm$  1.0 
\\ \tableline
         900    &                &                &                &
                &  3.7$\pm$  1.0 &  8.8$\pm$  1.1 &  7.7$\pm$  1.0 
\\ \tableline
        1100    &                &                &                &
                &                &  5.4$\pm$  0.9 &  7.8$\pm$  0.9 
\\ \tableline
        1300    &                &                &                &
                &                &  2.9$\pm$  0.9 &  6.7$\pm$  0.8 
\\ \tableline
        1500    &                &                &                &
                &                &                &  3.9$\pm$  0.8 
\\ 
\end{tabular}
\end{table}

\begin{table}
\caption{Invariant yields for $^{3}He$ in 38\% to 66\% most central
Au+Pb collisions in units of $10^{-3} c^{2}/GeV^{2}$.  Errors listed 
are systematic and statistical combined
in quadrature.  }
\label{tab:helium3_3866}
\begin{tabular}{|c|c|c|c|c|c|c|c|}
     Rapidity   &    1.0-1.2     &   1.2-1.4      &   1.4-1.6      &
  1.6-1.8       &    1.8-2.0     &   2.0-2.2      &   2.2-2.4     
\\ \tableline \tableline
$p_{T}(MeV/c)$  &                &                &                & 
                &                &                &                
\\ \tableline
         100    &  4.8$\pm$  2.3 &  1.9$\pm$  0.9 &  2.4$\pm$  1.0 &
                &
                &                &                
\\ \tableline
         300    &                &                &  2.5$\pm$  0.6 &
  1.9$\pm$  0.4 &
  2.3$\pm$  0.5 &  2.9$\pm$  0.6 &  7.0$\pm$  1.4 
\\ \tableline
         500    &                &                &                &
  1.5$\pm$  0.5 &
  2.5$\pm$  0.5 &  2.9$\pm$  0.4 &  5.2$\pm$  0.6 
\\ \tableline
         700    &                &                &                &
                &
  2.5$\pm$  0.6 &  2.2$\pm$  0.4 &  4.4$\pm$  0.5 
\\ \tableline
         900    &                &                &                &
                &
  1.4$\pm$  0.6 &  2.7$\pm$  0.5 &  3.6$\pm$  0.6 
\\ \tableline
        1100    &                &                &                &
                &
                &  1.7$\pm$  0.5 &  3.5$\pm$  0.6 
\\ \tableline
        1300    &                &                &                &
                &
                &                &  2.4$\pm$  0.5 
\\ \tableline
        1500    &                &                &                &
                &
                &                &  1.7$\pm$  0.4 
\\ 
\end{tabular}
\end{table}

\begin{table}
\caption{Invariant yields for tritons in 10\% most central
Au+Pb collisions in units of $10^{-3} c^{2}/GeV^{2}$.  Errors listed 
are systematic and statistical combined
in quadrature.  }
\label{tab:tritons}
\begin{tabular}{|c|c|c|c|c|c|c|c|}
     Rapidity &  1.0-1.2       &   1.2-1.4      &   1.4-1.6      
& 1.6-1.8        &
    1.8-2.0     &  2.0-2.2       &   2.2-2.4   \\ \tableline \tableline
$p_{T}$      &                &                &                
& 
                &                &             \\ \tableline
          50 & 12.5$\pm$  3.6 &                &                
&                &
                &                &                \\ \tableline
         150 & 16.0$\pm$  1.8 & 11.8$\pm$  1.5 &  9.8$\pm$  1.5 
&                &
                &                &                \\ \tableline
         250 & 13.6$\pm$  1.9 & 12.2$\pm$  1.5 &  9.9$\pm$  2.1 
& 11.8$\pm$  2.9 &
  9.3$\pm$  1.3 & 15.9$\pm$  2.5 &                \\ \tableline
         350 &                & 13.6$\pm$  1.7 &  9.6$\pm$  1.5 
& 12.2$\pm$  2.4 &
 10.4$\pm$  3.2 & 16.0$\pm$  1.8 & 16.1$\pm$  2.4 \\ \tableline
         450 &                &  9.2$\pm$  2.4 & 10.2$\pm$  1.5 
& 10.1$\pm$  1.7 &
 11.7$\pm$  2.5 & 16.0$\pm$  1.7 & 18.5$\pm$  2.1 \\ \tableline
         550 &                &                & 10.6$\pm$  1.4 
& 10.0$\pm$  1.5 &
 10.5$\pm$  3.1 & 15.8$\pm$  1.6 & 18.7$\pm$  1.9 \\ \tableline
         650 &                &                &                
&  9.0$\pm$  1.1 &
 11.4$\pm$  2.0 & 15.3$\pm$  1.7 & 17.5$\pm$  2.0 \\ \tableline
         750 &                &                &                
& 10.5$\pm$  1.5 &
  9.3$\pm$  1.6 & 14.8$\pm$  1.9 & 17.0$\pm$  1.9 \\ \tableline
         850 &                &                &                
&                &
 10.7$\pm$  1.5 & 16.7$\pm$  2.2 & 14.8$\pm$  2.0 \\ \tableline
         950 &                &                &                
&                &
 11.2$\pm$  1.3 & 14.1$\pm$  2.0 & 13.9$\pm$  2.5 \\ \tableline
        1050 &                &                &                
&                &
                & 12.9$\pm$  1.4 & 15.7$\pm$  1.8 \\ \tableline
        1150 &                &                &                
&                &
                &                & 12.2$\pm$  2.0 \\ \tableline
        1250 &                &                &                
&                &
                &                & 14.3$\pm$  1.9 \\ \tableline
        1350 &                &                &                
&                &
                &                &  9.1$\pm$  1.2 \\ \tableline
        1450 &                &                &                
&                &
                &                &  9.7$\pm$  1.3 \\ \tableline
\end{tabular}
\end{table}

\begin{table}
\caption{Invariant yields for $^{4}He$ in 10\% most central
Au+Pt collisions in units of $10^{-5} c^{2}/GeV^{2}$ 
from data taken in the 1998 run of E864.  
Errors listed are systematic and statistical combined
in quadrature.  }
\label{tab:helium4}
\begin{tabular}{|c|c|c|c|c|c|}
     Rapidity &  1.4-1.6       &   1.6-1.8      &   1.8-2.0      
& 2.0-2.2        &
    2.2-2.4   \\ \tableline \tableline
$p_{T}(MeV/c)$&                &                &                
& 
               \\ \tableline
         350 & 17.3$\pm$  1.9 & 17.4$\pm$  2.0 & 22.1$\pm$  2.6 
&                &
                \\ \tableline
         450 & 18.1$\pm$  2.4 & 16.7$\pm$  1.7 & 23.3$\pm$  2.5 
& 26.0$\pm$  3.1 &
                \\ \tableline
         550 & 17.4$\pm$  2.4 & 20.3$\pm$  2.6 & 19.4$\pm$  1.9 
& 28.0$\pm$  3.3 &
 43.7$\pm$ 11.4 \\ \tableline
         650 & 13.4$\pm$  2.1 & 18.0$\pm$  2.2 & 21.7$\pm$  2.3 
& 30.5$\pm$  3.5 &
 44.5$\pm$ 11.3 \\ \tableline
         750 & 20.4$\pm$  5.4 & 17.2$\pm$  2.3 & 21.1$\pm$  2.4 
& 29.8$\pm$  3.5 &
 45.6$\pm$ 11.4 \\ \tableline
         850 &                & 16.9$\pm$  2.3 & 21.2$\pm$  2.3 
& 31.7$\pm$  3.8 &
 47.1$\pm$ 11.8 \\ \tableline
         950 &                & 17.9$\pm$  2.8 & 20.3$\pm$  2.2 
& 28.9$\pm$  3.5 &
 45.8$\pm$ 11.4 \\ \tableline
        1050 &                & 15.5$\pm$  3.0 & 23.6$\pm$  2.8 
& 27.6$\pm$  3.5 &
 42.3$\pm$ 10.6 \\ \tableline
        1150 &                &                & 19.8$\pm$  2.3 
& 27.1$\pm$  3.4 &
 38.5$\pm$  9.7 \\ \tableline
        1250 &                &                & 18.7$\pm$  2.3 
& 27.6$\pm$  3.4 &
 41.6$\pm$ 10.6 \\ \tableline
        1350 &                &                & 23.4$\pm$  4.0 
& 23.9$\pm$  3.0 &
 37.0$\pm$  9.5 \\ \tableline
        1450 &                &                &                
& 25.8$\pm$  3.3 &
 32.0$\pm$  8.4 \\ \tableline
        1550 &                &                &                
& 21.1$\pm$  2.8 &
 30.7$\pm$  8.1 \\ \tableline
        1650 &                &                &                
& 22.3$\pm$  3.3 &
 29.1$\pm$  7.6 \\ \tableline
        1750 &                &                &                
& 20.8$\pm$  3.5 &
 21.5$\pm$  5.8 \\ \tableline
        1850 &                &                &                
& 15.3$\pm$  2.7 &
 21.2$\pm$  5.7 \\ \tableline
        1950 &                &                &                
&                &
 18.5$\pm$  5.1 \\ \tableline
        2050 &                &                &                
&                &
 13.4$\pm$  3.9 \\ \tableline
        2150 &                &                &                
&                &
 16.5$\pm$  5.2 \\ \tableline
\end{tabular}
\end{table}

\begin{table}
\caption{Invariant yields for $^{6}He$ and $^{6}Li$ in Au+Pt
collisions in units of
$10^{-8} c^{2}/GeV^{2}$.  Errors listed are systematic 
and statistical combined in quadrature.  }
\label{tab:mass6}
\begin{tabular}{|c|c|c|}
($y,p_{T}(MeV/c)$)      &  $^{6}He$      & $^{6}Li$       
\\ \tableline \tableline
(1.6-1.8,500-1000)      &  6.0 $\pm$ 1.8 &                \\ \tableline
(1.8-2.0,0-500)         & 10.5 $\pm$ 4.6 &                \\ \tableline
(1.8-2.0,500-1000)      &  5.2 $\pm$ 1.2 & 32 $\pm$ 15    \\ \tableline
(1.8-2.0,1000-1500)     &  3.3 $\pm$ 1.2 & 24.9 $\pm$ 6.1 \\ \tableline
(2.0-2.2,500-1000)      & 13.8 $\pm$ 1.2 & 33.7 $\pm$ 6.9 \\ \tableline
(2.0-2.2,1000-1500)     & 10.4 $\pm$ 1.2 & 37.5 $\pm$ 7.8 \\ \tableline
(2.0-2.2,1500-2000)     &                & 34 $\pm$ 21    \\
\end{tabular}
\end{table}

\begin{table}
\caption{Invariant yields for $^{7}Li$ and $^{7}Be$ in Au+Pt 
collisions in units of
$10^{-8} c^{2}/GeV^{2}$.  Errors listed are systematic 
and statistical combined in quadrature.}
\label{tab:mass7}
\begin{tabular}{|c|c|c|}
($y,p_{T}(GeV/c)$)        &  $^{7}Li$      & $^{7}Be$              
\\ \tableline \tableline
(1.6-2.2,0-1.8)    &  .92  $\pm$ .4 &                       \\ \tableline
(2.0-2.2,0-1.8)    &                & 1.3 $\pm ^{.35}_{.25}$\\
\end{tabular}
\end{table}


\begin{references}
\bibitem[\ast]{}        Present address: Vanderbilt University, 
Nashville, Tennessee 37235 
\bibitem[\dag]{}        Present Address: Istituto di Cosmo-Geofisica del CNR,
Torino, Italy / INFN Torino, Italy
\bibitem[\ddag]{}       Present Address: Anderson Consulting, Hartford, CT
\bibitem[\S]{} 	Present address: Univ. of Denver, Denver CO 80208
\bibitem[\|]{}    	Deceased.
\bibitem[\P]{}     	Present address: Cambridge Systematics, 
Cambridge, MA 02139
\bibitem[\ast\ast]{}    Present address: McKinsey \& Co., New York, NY 10022
\bibitem[\dag\dag]{}    Present address: Department of Radiation Oncology, 
Medical College of Virginia, Richmond VA 23298
\bibitem[\ddag\ddag]{}  Present address: University of Tennessee, 
Knoxville TN 37996
\bibitem[\S\S]{}        Present address: Institut de Physique 
Nucl\'{e}aire, 91406 Orsay Cedex, France
\bibitem[\|\|]{} 	Present Address: Institute for Defense 
Analysis, Alexandria VA 22311
\bibitem[\P\P]{} 	Present Address: MIT Lincoln Laboratory, 
Lexington MA 02420-9185

\bibitem{pratt}S. Pratt, Phys. Rev. Lett. 53(1984)1219.
\bibitem{dq1and2} L. C. Alexa et. al., Phys. Rev. Lett. 82{(1999)} 1374.
D. Abbott et al., Phys. Rev. Lett. 82{(1999)} 1379.
\bibitem{sjohn} S.C. Johnson, PhD Thesis, SUNY Stony Brook 1997.
\bibitem{sato} H. Sato and K. Yazaki, Phys. Lett. B 98{(1981)} 153.
\bibitem{bond} R. Bond, P.J. Johansen, S.E. Koonin, and S. Garpman, 
 Phys. Lett. B 72(1981) 131.
\bibitem{nagle} J.L. Nagle, B.S. Kumar, D. Kusnezov ,
H. Sorge and R. Matiello, Phys. Rev. C 53(1996) 367.
\bibitem{na52ba} G. Ambrosini et. al., Nucl. Phys. A610(1996) 306.
\bibitem{pbm} P. Braun-Munzinger, J. Stachel, J.P. Wessels, and N. Xu, 
 Phys. Lett. B 344(1995) 43.
\bibitem{uliflow} H.Dobler, J.Sollfrank, and U. Heinz 
 Phys. Lett. B 457(1999) 353.
\bibitem{RQMD} H. Sorge, A. von Keitz, R. Matiello, H. Stocker,
and W. Greiner, Nucl. Phys. A525(1991) 95c.
\bibitem{heinzprc}R. Scheibl and U. Heinz, Phys. Rev. C. 59(1999)1585.
\bibitem{polleri} A. Polleri, J.P. Bondorf, and I.N. Mishustin,
Phys. Lett. B 419(1998) 19.
\bibitem{mek} S. Das Gupta and A.Z. Mekjian, Phys. Rep. C 53(1996) 367.
\bibitem{pbmtherm} P. Braun-Munzinger, J. Stachel, J.P. Wessels, and N. Xu, 
 Phys. Lett. B 365(1996) 1.
\bibitem{nag81} S. Nagamiya et. al., Phys. Rev. C 24(1981) 971.
\bibitem{bennet}M.J. Bennett et. al., Phys. Rev. C. 58(1998)1155.
\bibitem{abb94} T. Abbott et. al., Phys. Rev. C 50{(1994)} 1024.
\bibitem{bignim}T. A. Armstrong et al., Nucl. Inst. Meth. A437(1999)222.
\bibitem{beam} P. Haridas, I.A. Pless. G. Van Buren, J. Tomasi, 
M.S.Z. Rabin, K. Barish and R.D. Majka, Nucl. Inst. Meth. A385(1997) 413.
\bibitem{alexi} A. Chikanian, B.S. Kumar, N. Smirnov and E.O'Brien, 
Nucl. Inst. Meth. A371(1996) 480.
\bibitem{calo}T.A. Armstrong et. al., Nucl. Inst. Meth. A406(1998)227.
\bibitem{let}J.C. Hill et. al., Nucl. Inst. Meth. A421(1999)431.
\bibitem{neut} T.A. Armstrong et. al., Phys. Rev. C 60{(1999)}064903.
\bibitem{theses}Ph.D. theses: N.K. George, Z. Xu, S.D. Coe, J.K. Pope, 
L.E. Finch (Yale University) ,R. Hoversten (Iowa State University, 
in preparation).
\bibitem{e891} S. Ahmad et. al., Phys. Lett. B 382(1996) 35.
\bibitem{e877} J. Barrette et. al., nucl-ex/9906005.
\bibitem{nkg_parkcity} N.K. George, Talk presented at Park City Utah, 
January 9-16, 1999. 
Proceedings to be published by Kluwer Academic Press, 1999.
\bibitem{pbarprc} T.A. Armstrong et. al., Phys. Rev. C 59{(1999)} 2699.
\bibitem{llope} W.J. Llope et.al., Phys. Rev. C 52{(1995)} 2004.
\bibitem{mm} M. Murray for the NA44 Collaboration, 
ICPAQGP 97 Conference.; nucl-ex/9706007.
\bibitem{mati}R. Mattiello, H.Sorge, H.Stocker, and W. Griener, 
 Phys. Rev. C. 55(1997)1443.
\bibitem{eos}S. Wang et. al., Phys. Rev. Lett. 74{(1995)} 2646.
\bibitem{xzb_ismd} Z. Xu for the E864 Collaboration, 
ISMD 99 Conference.; nucl-ex/9909012.
\bibitem{xzbprl} T.A. Armstrong et al., Phys. Rev. Lett. 83{(1999)} 5431.

\end{references}
\end{document}